\documentclass[sigconf]{acmart}
\usepackage{graphicx}
\usepackage{amsmath}
\usepackage{booktabs}
\usepackage{amsfonts}
\usepackage{multirow}[c]
\usepackage{makecell}
\usepackage{subfig}
\usepackage{color}
\usepackage{bm}
\usepackage{epstopdf}
\usepackage{url}
\usepackage[cal=cm]{mathalfa}
\usepackage{balance}
\usepackage{threeparttable}
\usepackage{wrapfig}
\usepackage{tabularx}
\usepackage{array}
\usepackage{enumitem}
\usepackage{appendix}
\usepackage{tikz}
\usepackage{float} 
\usepackage{adjustbox}
\usepackage[linesnumbered,ruled,vlined]{algorithm2e}

\SetKwProg{Stage}{Stage}{}{} 

\setlist[itemize]{leftmargin=*}

\author{Ruochen Yang}
\authornote{Equal contribution.}
\affiliation{
  \institution{Institute of Information Engineering, Chinese Academy of Sciences  \\ School of Cyber Security, UCAS$^\ddag$}
  \city{Beijing}
  \country{China}
}
\email{yangruochen@iie.ac.cn}

\author{Xiaodong Li}
\authornotemark[1]
\affiliation{
  \institution{Institute of Information Engineering, Chinese Academy of Sciences  \\ School of Cyber Security, UCAS}
  \city{Beijing}
  \country{China}
}
\email{lixiaodong@iie.ac.cn}

\author{Jiawei Sheng}
\affiliation{
  \institution{Institute of Information Engineering, Chinese Academy of Sciences}
  \city{Beijing}
  \country{China}
}
\email{shengjiawei@iie.ac.cn}

\author{Jiangxia Cao}
\affiliation{
  \institution{Kuaishou Technology}
  \city{Beijing}
  \country{China} 
}
\email{caojiangxia@kuaishou.com}

\author{Xinkui Lin}
\affiliation{
  \institution{Institute of Information Engineering, Chinese Academy of Sciences  \\ School of Cyber Security, UCAS}
  \city{Beijing}
  \country{China} 
}
\email{linxinkui@iie.ac.cn}

\author{Shen Wang}
\affiliation{
  \institution{Kuaishou Technology}
  \city{Beijing}
  \country{China} 
}
\email{wangshen@kuaishou.com}

\author{Shuang Yang}
\affiliation{
  \institution{Kuaishou Technology}
  \city{Beijing}
  \country{China}   
}
\email{yangshuang08@kuaishou.com}

\author{Zhaojie Liu}
\affiliation{
  \institution{Kuaishou Technology}
  \city{Beijing}
  \country{China}  
}
\email{zhaotianxing@kuaishou.com}

\author{Tingwen Liu}
\authornote{Corresponding author. \\ $^\ddag$ University of Chinese Academy of Sciences}
\affiliation{
  \institution{Institute of Information Engineering, Chinese Academy of Sciences  \\ School of Cyber Security, UCAS}
  \city{Beijing}
  \country{China}  
}
\email{liutingwen@iie.ac.cn}

\AtBeginDocument{}

\copyrightyear{2026}
\acmYear{2026}
\setcopyright{cc}
\setcctype{by}
\acmConference[KDD '26]{Proceedings of the 32nd ACM SIGKDD Conference on Knowledge Discovery and Data Mining V.1}{August 09--13, 2026}{Jeju Island, Republic of Korea}
\acmBooktitle{Proceedings of the 32nd ACM SIGKDD Conference on Knowledge Discovery and Data Mining V.1 (KDD '26), August 09--13, 2026, Jeju Island, Republic of Korea}
\acmPrice{}
\acmDOI{10.1145/3770854.3780285}
\acmISBN{979-8-4007-2258-5/2026/08}

\title{From Agnostic to Specific: Latent Preference Diffusion for Multi-Behavior Sequential Recommendation}

\begin{document}

\begin{abstract}

Multi-behavior sequential recommendation (MBSR) aims to learn the dynamic and heterogeneous interactions of users' multi-behavior sequences, so as to capture user preferences under target behavior for the next interacted item prediction. 
Unlike previous methods that adopt unidirectional modeling by mapping auxiliary behaviors to target behavior, recent concerns are shifting from behavior-fixed to behavior-specific recommendation.
However, these methods still ignore the user's latent preference that underlying decision-making, leading to suboptimal solutions.
Meanwhile, due to the asymmetric deterministic between items and behaviors, discriminative paradigm based on preference scoring is unsuitable to capture the uncertainty from low-entropy behaviors to high-entropy items, failing to provide efficient and diverse recommendation.
To address these challenges, we propose \textbf{FatsMB}, a framework based diffusion model that guides preference generation \textit{\textbf{F}rom Behavior-\textbf{A}gnostic \textbf{T}o Behavior-\textbf{S}pecific} in latent spaces, enabling diverse and accurate \textit{\textbf{M}ulti-\textbf{B}ehavior Sequential Recommendation}. Specifically, we design a Multi-Behavior AutoEncoder (MBAE) to construct a unified user latent preference space, facilitating interaction and collaboration across Behaviors, within Behavior-aware RoPE (BaRoPE) employed for multiple information fusion. Subsequently, we conduct target behavior-specific preference transfer in the latent space, enriching with informative priors. A Multi-Condition Guided Layer Normalization (MCGLN) is introduced for the denoising. Extensive experiments on real-world datasets demonstrate the effectiveness of our model. 
We release our code at \url{https://github.com/OrchidViolet/FatsMB}.

\end{abstract}

\ccsdesc[500]{Information systems~Recommender systems}

\keywords{Multi-behavior Sequential Recommender, Latent Diffusion Model}

\maketitle

\section{Introduction} \label{sec:intro}

Recommendation System (RS) aims to deliver items of the best-matched interest to users based on their historical interactions, which has been widely deployed across various platforms~\cite{din, moment, farm}. In particular, to capture the temporal evolution of user preferences, Sequential Recommendation (SR)~\cite{sasrec, gru4rec,  bert4rec} introduces interaction sequences information to achieve dynamic interest modeling over time. However, from a more realistic perspective, user interactions are not limited to a single type of behavior~\cite{nmtr, matn}. Instead, heterogeneous actions, such as \textit{browsing}, \textit{liking} and \textit{purchasing}, are interdependent and exhibit complex causal couplings. Therefore, the way to integrate multi-behavior interaction information into sequential recommendation has become an increasing attention.

Compared to single-behavior, multi-behavior sequential recommendation (MBSR) faces two primary challenges. The first lies in the fusion of heterogeneous information. Obviously, different behaviors reflect distinct aspects of user preferences, which emphasizes the ability for capturing complex dependency patterns. 
Previous methods~\cite{mbgcn, mb-gmn, mbht} conduct separate learning for subsequences of each behavior type and then combine into a unified representation through adaptive weighting aggregation, focusing only on the dependencies between behaviors.
Alternatively,~\cite{mb-str, ghtid, m-gpt} prioritize capturing item dependencies within sequences, aiming to characterize the evolution of user intent as it propagates across interacted items over time. However, most of them require a fixed target behavior, means the prediction under a certain given behavior, which limits the applicability to a single recommendation scenario.

\begin{figure}[t!]
\begin{center}
\includegraphics[width=8.5cm]{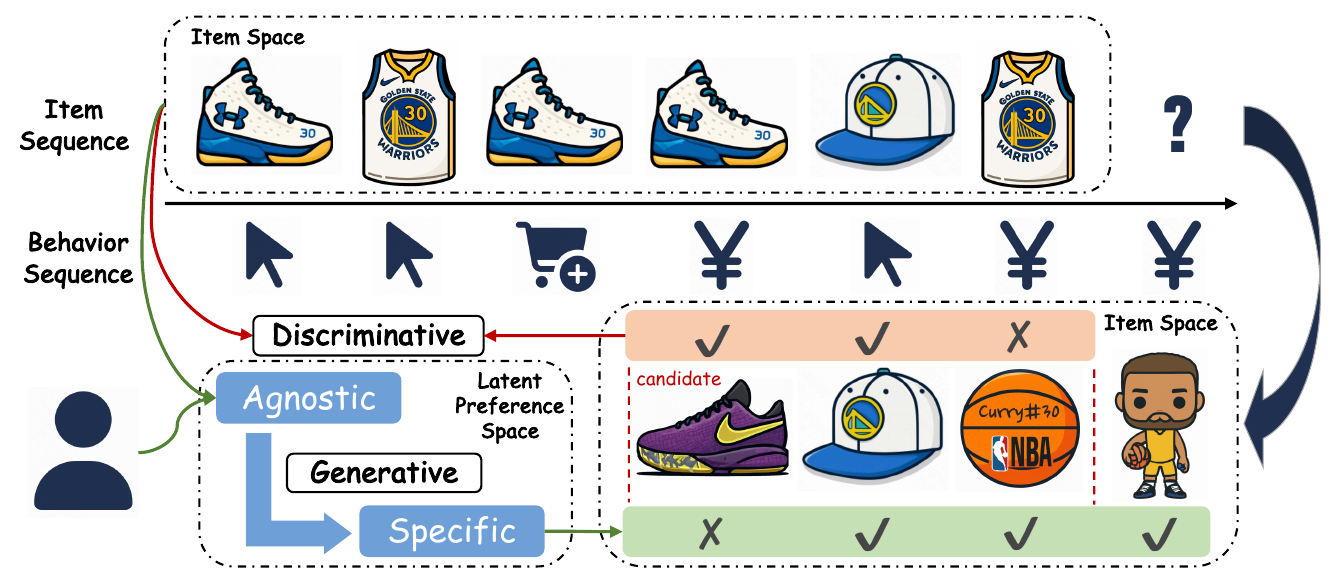}
\vspace{-0.6cm}
\caption{Motivation of our work. Discriminative paradigm is constrained by candidate set and interaction history, while generative paradigm incorporating user's latent preference achieves accuracy and diversity.}
\vspace{-0.4cm}
\label{fig:motivation}
\end{center}
\end{figure}

Similarly, some perspectives argue that heterogeneous behaviors are equally important in meeting user demands in multiple aspects~\cite{bvae}, while the single behavior recommendation ignores the objective needs of both users and platforms. Therefore, another challenge lies in the transition from behavior-fixed to behavior-specific recommendation, that is, producing suggestion lists across multiple behavior tasks. Recent methods~\cite{mb-str, pbat, missl} have attempted to address this problem by either leveraging knowledge transfer cross behaviors or modeling the hierarchical progression of behaviors. Although these methods have shown promising effectivity, they still ignore the user's latent preference underlying the decision-making, which is detached from specific behavior and exhibits behavior-agnostic characteristic. As illustrated in Figure \ref{fig:motivation}, focusing more on purchase history mistakenly identify the user as a sports equipment enthusiast, while overlooking the user's inherent motivation behind his interactions is the fandom for \textit{Curry}. This perceptual bias caused by behavior may lead to suboptimal solutions.

Note that the aforementioned methods mostly adopt discriminative modeling paradigm, which means the learning for conditional probability given user-item pair with the certain behavior. Although such methods have achieved significant results, they suffer from fundamental limitations: discriminative paradigm treats recommendation as a series of independent classification tasks, which not only confines predictions in the candidate sets but also hinders the generation of diversity~\cite{dreamrec}. 
In fact, MBSR requires predicting the items that users likely to interact with under a specific behavior, rather than judging whether users perform certain behaviors on a given item. This inversion of the causal direction is a critical bottleneck of discriminative paradigm, preventing the model aligns with user intentions in large scale scenarios.

Furthermore, from an information theory perspective, we find that there exists an asymmetric deterministic between items and behaviors. In brief, user behavior toward a given item is relatively predictable, whereas the specific items associated with a given behavior exhibit considerable uncertainty. This intuitive observation can also be evident in the substantial discrepancy between bidirectional cross entropy values in Table \ref{tab:entropy}. 
Reconstructing high-entropy items from low-entropy behaviors is not suitable for discriminative paradigm due to its difficulty in modeling this uncertainty.
Therefore, we argue that behavior-specific recommendation tasks should be addressed using generative models, as they can effectively model the latent preference that drive user behaviors.

To address the above problems, it's necessary to leverage the user's inherent intentions, as the intuitive way is generating based on behavior-agnostic preferences. 
However, the item space constructed from users' interaction sequences couples various behavior information, making it difficult to disentangle the latent preferences and perform target matching. Inspired by the effectiveness of latent diffusion models (LDM)~\cite{ldm} in transfer tasks~\cite{mcm-ldm, styleid}, for one hand they utilize the low-dimensional, smooth and continuous characteristics of latent space~\cite{oldphoto} for complex pattern capture and separation, and for the other hand, their powerful guided generation capability is suitable for acquiring target behavior-specific preferences. 
Therefore, we propose \textbf{FatsMB} based on LDM, which centers on the transfer \textbf{F}rom Behavior-\textbf{A}gnostic \textbf{T}o Behavior-\textbf{S}pecific in the latent space. Concretely speaking, user interaction sequences are encoded into the unified latent preference space through Multi-Behavior AutoEncoder (MBAE), where the encoder utilizes Behavior-aware RoPE (BaRoPE) to address the multi-perspective capture of multi-behavior information. In the latent space, a diffusion model is employed to transfer preference representations from behavior-agnostic to behavior-specific, where Multi-Condition Guided Layer Norm (MCGLN) serves as denoiser. Finally, the target behavior-specific representations are decoded back into the item space to generate recommended items. 

\begin{table}
\setlength{\abovecaptionskip}{2pt}
\caption{The \textit{entropy} $\text{H}(\cdot)$, \textit{conditional entropy} $\text{H}(\cdot|\cdot)$ and \textit{mutual information} $\text{I}(\cdot, \cdot)$ between \textit{items I} and \textit{behaviors B} of the three datasets. Details refer to Appendix.\ref{app:entropy}.}
\vspace{+0.2cm}
\label{tab:entropy}
\centering
\begin{small}
  \begin{tabular}{c|c|c|c|c|c}
    \toprule
    \textbf{Dataset} & \textbf{$\text{H}(I)$} & \textbf{$\text{H}(B)$} & \textbf{$\text{H}(B|I)$} & \textbf{$\text{H}(I|B)$} & \textbf{$\text{I}(I,B)$}\\
    \midrule
    Yelp & 9.6591 & 1.2540 & 1.1569 & 9.5620 & 0.0971 \\
    Retail & 10.8756 & 0.7227 & 0.6702 & 10.8231 & 0.0525\\
    IJCAI & 12.1743 & 0.5691 & 0.5041 & 12.1093 & 0.0651\\
    \bottomrule 
  \end{tabular}
\end{small}
\vspace{-0.2cm}
\end{table}

The main contributions of this paper are summarized as follows:
\begin{itemize}
    \item We identify the neglect of latent preferences and limitations of discriminative paradigms as key causes of suboptimal performance in behavior-specific MBSR, and propose a novel method based on the transfer and generation of preferences for this task.
    \item We design FatsMB base on latent diffusion and associated components for behavior-specific MBSR. To the best of our knowledge, this is the first attempt to introduce LDM in this domain.
    \item We conduct extensive experiments on publicly available datasets and demonstrate the effectiveness of our model and components.
\end{itemize}

\section{Preliminary}

\subsection{Problem Formulation}

Similarly to the typical MBSR scenario, we define the set of users as $\mathcal{U} = \{u_1,u_2,\dots,u_{\lvert \mathcal{U} \rvert} \}$ and items as $\mathcal{V} = \{v_1,v_2,\dots,v_{\lvert \mathcal{V} \rvert} \}$, where $\lvert \mathcal{U} \rvert$ and $\lvert \mathcal{V} \rvert$ denote the number of users and items. Then we formulate the set of behaviors as $\mathcal{B} = \{b_1,b_2,\dots,b_{\lvert \mathcal{B
} \rvert} \}$, supposing there are $\lvert \mathcal{B
} \rvert$ types of behaviors. For an individual user $u \in \mathcal{U}$, whose multi-behavior interaction sequence could be written as $\mathcal{S}_u = \{ \langle v^u_1, b^u_1 \rangle, \langle v^u_2, b^u_2 \rangle,\dots,\langle v^u_{\lvert \mathcal{S}_u \rvert}, b^u_{\lvert \mathcal{S}_u \rvert} \rangle \}$, where $\lvert \mathcal{S}_u \rvert$ is the length of the sequence, we render $\lvert \mathcal{S}_u \rvert = L$ for all sequences for a fixed length by padding the insufficient part with special token [padding], and $\langle v^u_i, b^u_i \rangle$ is the $i$-th behavior-aware user-item interaction pair between user $u \in \mathcal{U}$ and item $v^u_i \in \mathcal{V}$ under behavior $b^u_i \in \mathcal{B}$.
In the previous MBSR task, the target behavior is typically fixed, but we argue that the item prediction across different behaviors is of equal importance, thus we aim to perform item prediction for any given behavior instead of fixing the target behavior.

\subsection{Diffusion Model}

\textbf{Diffusion Models (DM)}~\cite{ddpm,improved-ddpm, ddim} typically consist of two diffusion process, \textit{i.e}., a forward process and a reverse process. The forward process gradually adds noise to the input image $x_0$ and producing a series of intermediate states $x_1,\dots,x_T$ at each timestep $t\in[1,T]$. This process can be defined as:
\begin{equation}
\label{forward_process}
    q(x_t|x_{t-1}) := \mathcal{N}(x_t; \sqrt{1-\beta_t}x_{t-1}, \beta_t\mathbf{I}),
\end{equation}
where $\beta_t\in(0,1)$ is a fixed variance schedule. 
Besides, the reverse process of \textbf{Denoising Diffusion Probabilistic Model (DDPM)}~\cite{ddpm, improved-ddpm} aims to reconstruct the initial representation $x_0$ through $T$ steps of denoising from a Gaussian noise sample $x_T\sim\mathcal{N}(0,\mathbf{I})$, formulated as:
\begin{equation}
\label{reverse_process}
    p_{\theta}(x_{t-1} | x_t) := \mathcal{N}(x_{t-1}; \mu_{\theta}(x_t, t), \sigma_t^2\mathbf{I}),    
\end{equation}
where $\mu_{\theta}(x_t, t)$ denotes the predicted mean of the approximate distribution, and $\sigma_t$ represents the variance that controls the magnitude of randomness.
Therefore, following the Markov chain rule, the reverse process can be expressed as:
\begin{equation}
    x_{t-1} := \frac{1}{\sqrt{\alpha_t}} \left( x_t - \frac{1 - \alpha_t}{\sqrt{1 - \bar{\alpha}_t}} \epsilon_\theta(x_t, t) \right) + \sigma_t \epsilon,
\end{equation}
where $\alpha_t:=1-\beta_t$, $\bar{\alpha}_t := \prod_{s=0}^t \alpha_s.$, $\epsilon\sim\mathcal{N}(0,\mathbf{I})$, and $\epsilon_\theta(x_t, t)$ is the noise predicted by the neural networks parameterized by $\theta$. The corresponding objective can be simplified to:
\begin{equation}
    L_{DM} := \mathbb{E}_{x, \epsilon \sim \mathcal{N}(0,\mathbf{I}), t} \left[ \| \epsilon - \epsilon_\theta(x_t, t) \|^2 \right].
\end{equation}

\noindent \textbf{Denoising Diffusion Implicit Model (DDIM)}~\cite{ddim} removes the Markovian assumption of the diffusion process, allowing for non-consecutive time jumps. This enables an accelerated generation process and results in deterministic sampling:
\begin{equation}
    x_{t-1} := \sqrt{\frac{\alpha_{t-1}}{\alpha_t}} (x_t - \sqrt{1 - \alpha_t} \epsilon_\theta(x_t,t)) + \sqrt{1 - \alpha_{t-1}} \epsilon_\theta(x_t,t).
\end{equation}

\noindent \textbf{Latent Diffusion Model (LDM)}~\cite{ldm} transfers the diffusion model from the high dimensional space of pixel to the low dimensional latent space of representation using a Variational Autoencoder (VAE)~\cite{vae} with encoder $\mathcal{E}$ and decoder $\mathcal{D}$, for focusing on semantic bits of data and reduce computation costs. The diffusion and denoising process are both performed in the compressed latent space defined by the encoder, iteratively reconstructing a sample from random noise based on conditional information $y$ and then refining the details through the decoder. The objective is now modified to the following:
\begin{equation}
\label{objective_ldm}
    L_{LDM} := \mathbb{E}_{\mathcal{E}(x), \epsilon \sim \mathcal{N}(0,\mathbf{I}), t} \left[ \| \epsilon - \epsilon_\theta(z_t, t, y) \|^2 \right].
\end{equation}

Considering the transformation from the user interest space to the behavior decision space aligns well with the guided generation in latent space, which is subsequently reconstructed by the decoder into preferred items. Therefore, applying \textbf{Latent Diffusion Model} to \textbf{Multi-Behavior Sequential Recommendation} presents an intuitive yet effective approach.

\section{Methodology}

\begin{figure}[t!]
\vspace{-0.1cm}
\begin{center}
\includegraphics[width=8.5cm]{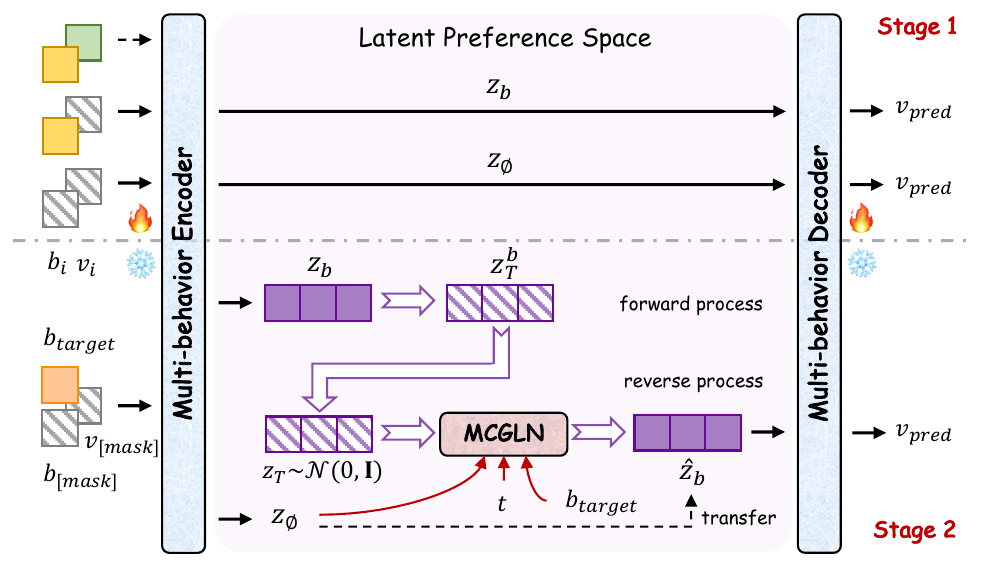}
\vspace{-0.5cm}
\caption{The overall framework of our model. Stage 1 trains MBAE to construct the latent space for unifying preferences. Stage 2 achieves the transfer from behavior-agnostic to behavior-specific in the latent space based on DM.}
\vspace{-0.5cm}
\label{fig:model}
\end{center}
\end{figure}

In this section, we introduce our proposed framework in details, as depicted in Figure \ref{fig:model}.
Central to our method is (i) a joint latent space Multi-behavior autoencoder (Section \ref{sec:mbae}) that encodes behavior preferences from multi-behavior interaction sequences and decodes to the target item; and (ii) a continuous diffusion models FatsMB (Section \ref{sec:fatsmb}) enables the transfer of preferences from behavior-agnostic to behavior-specific, during which we propose a novel Multi-condition Guided Denoiser (Section \ref{sec:mgcln}) in conjunction with multiple constraint information for the denoising process.

\subsection{Multi-Behavior AutoEncoder} \label{sec:mbae}

Following common practice~\cite{ldm, mcm-ldm, planner, simdiff}, we do not directly apply the generative model to the raw data. 
Instead, we first encode the preferences of a given user interaction sequence under different target behaviors into a unified low-dimensional space.
This latent space constructed by the pre-trained autoencoder is the coupling of all behavior preferences, and the diffusion model is consequently trained within the latent space for delicate transfer and refinement.  

In previous LDM works~\cite{ldm, styleid, planner}, in order to avoid arbitrary variations in the latent space, a slight KL-penalty is commonly applied to achieve a certain degree of spatial constraint and compression, similar to the KL divergence in VAE~\cite{vae}: $KL(p_\phi(z|x)||\mathcal{N}(0,\mathbf{I}))$, where $z = \mathcal{E}(x)$ is the encoded representation of the input. 
However,~\cite{va-vae} shows that a more uniform distribution of feature representations in the latent space weakens reconstruction capability. 
Considering our model aims to derive target behavior preferences from users' unified preferences rather than guided generating through complete sampling from the constrained distribution, which emphasizing the reconstruction capability of target behavior preferences (Section \ref{section_training}). 
Therefore, to better preserve the detailed information of multi-behavior sequences and avoid excessive compression, we employ an autoencoder (AE)~\cite{ae} architecture, named \textbf{Multi-Behavior AutoEncoder (MBAE)}, which discards the KL-penalty and spatial sampling compared with VAE~\cite{contrastvae, bvae}.

Since the quality of the encoder determines the upper bound of performance for downstream tasks, 
it's important to apply an architecture capable of capturing sequential information for adapting to the task of encoding from sequence to unified behavior space during the pre-training stage.
Note that BERT~\cite{bert} based on self-attention has achieved great success in discrete text sequence understanding tasks, and subsequently BERT4Rec~\cite{bert4rec} leveraging bidirectional modeling has demonstrated the aggregation of context facilitates sequence representation learning in sequential recommendation task.
Therefore, we employ a similar Transformer layers~\cite{attention} that utilize multi-head self-attention mechanisms and stack $l$ layers to serve as our encoder architecture:
\begin{equation}
    \text{MultiHead}(Q,K,V) = \text{Concat}(head_1,\dots,head_h)W^O,
\end{equation}
\begin{equation}
    where \quad head_i = \text{softmax}(\frac{((QW_i^Q)(KW_i^K))^T}{\sqrt{d_k}})(VW_i^V),
\end{equation}
where $W_i^Q,W_i^K,W_i^V \in \mathbb{R}^{d \times d_k}$ and $W^O \in \mathbb{R}^{hd_k \times d}$ are the corresponding projection matrices. $h$ is the number of heads, and $d = hd_k$ is the dimension of embeddings. 

In self-attention, $Q=K=V$ are the hidden input representations at each Transformer layer, as the embeddings of inputs for the first layer. For MBSR scenario, the initial embedding of multi-behavior sequence is usually obtained through a behavior-aware sequence embedding layer~\cite{missl, end4rec}, which is the integration of the triplet tokens $\langle v,b,p \rangle$ corresponding to each position in the sequence:
\begin{equation}
\label{absolute_position}
    h_i = e_{v_i} + e_{b_i} + e_{p_i}, \quad H = [h_0,h_1,...,h_L]
\end{equation}
where $e_{v_i},e_{b_i},e_{p_i} \in \mathbb{R}^d$ is the embeddings of the item, behavior and position, and $H \in \mathbb{R}^{L \times d}$ is the behavior-aware embedding matrix for the sequence of fixed length $L$. 

However, users' multi-behavior interactions are typically position-sensitive, while absolute positional encoding like Eq.\ref{absolute_position} fails to capture this positional sensitivity, making it ineffective in modeling the differences between distinct sequences~\cite{transformer-xl}. Inspired by the Rotary Position Embedding (RoPE)~\cite{roformer} that widely used in large language models~\cite{llama-2, qwen2}, which can effectively capture relative positional information and retain absolute characteristic. We propose \textbf{Behavior-aware RoPE (BaRoPE)}, to achieve the fusion of items, behaviors and positions within the sequence. 

\begin{figure}[H]
\vspace{-0.2cm}
\begin{center}
\includegraphics[width=5cm]{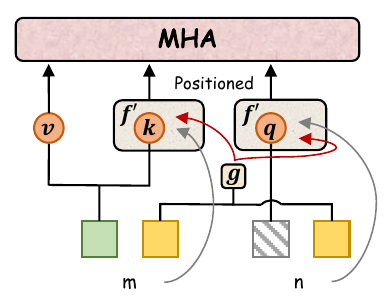}
\vspace{-0.3cm}
\caption{The architecture of BaRoPE.}
\vspace{-0.3cm}
\label{fig:arc_barope}
\end{center}
\end{figure}

In simple terms, RoPE aims to find an appropriate method $\tilde{q_m}=f(q,m)$ and $\tilde{k_n}=f(k,n)$ for incorporating absolute positional information $m,n$ to achieve the relative positional relationship, which is achieved through the multiplication in the complex number field:
\begin{equation}
    \langle f(q,m), f(k,n) \rangle = g(q,k,m-n),
\end{equation}
\begin{equation}
\label{eq:rope}
    where \quad f(q,m) = qe^{im\theta}, \quad f(k,n)=ke^{in\theta}.
\end{equation}
For applying to multi-behavior sequence, each token incorporates behavior information $a,b$ extra while maintaining the objective:
\begin{equation}
\label{BaRoPE_formula}
    \langle f'(q,a,m), f'(k,b,n) \rangle = g'(q,k,a,b,m-n).
\end{equation}
Based on the asymmetric deterministic between items and behaviors (Section \ref{sec:intro}), the determination of items is fundamentally accompanied by the determination of user behaviors, this means the introduction of behavior information doesn't affect the item information at the corresponding position. Therefore, we can approximately decouple the behavior information from the position information encoding:
\begin{equation}
\label{eq:barope}
    f'(q,a,m) = h(a)f(q,m), \quad f'(k,b,n) = h(b)f(k,n),
\end{equation}
$h(\cdot)$ is a fit for behavior representation, and we implement it using MLP. On this basis, the Eq.\ref{BaRoPE_formula} is established accordingly. 
The usage of BaRoPE in multi-head attention is depicted in Figure \ref{fig:arc_barope}, as it only affects $q$ and $v$, adding behavior and position information to the computation of attention score. Detailed derivation in Appendix.\ref{app:derivation}.

For the training of the encoder, we adopt the Cloze task~\cite{bert, bert4rec}, where each token in the sequence is masked with a special token [mask] at a certain probability $\rho$, and utilizing the context information for reconstruction. Specifically, after processing the sequence through Transformer layers with BaRoPE, we obtain the final output representation $z_b = h_{b}^{l}$ at the masked position, which we refer to as the user's behavior-specific preference of $b$. The decoder aims at recovering the masked item $v_{[mask]}$ utilizing this representation, also means the transfer from preference space to item space, which we implement MLP to achieve. Therefore, the objective of MBAE is unifying multi-behavior preferences and reconstructing the masked item:
\begin{equation}
\label{objective_mbae}
    L_{MBAE} := \mathbb{E}_{p_{\phi}(z_b|S_u)}[\text{log}  q_\psi(v_{[mask]}|z_b)].
\end{equation}

Furthermore, to learn consistent unified behavior preference, that is, to unify the behavior-agnostic and behavior-specific preference within the same latent preference space, we further mask the behavior corresponding to the masked item position with a certain probability $\sigma$. 
The final representation $z_\varnothing = h^l_{[mask]}$ encoded after masking the behavior information is considered the user's behavior-agnostic preference representation, since it abandon the target information and focus entirely on interactions to obtain the overall perception. And this representation serves as the transfer subject for the subsequent LDM.

\subsection{Latent Preference Diffusion} \label{sec:fatsmb}

\begin{figure}[t!]
\vspace{-0.1cm}
\begin{center}
\includegraphics[width=8cm]{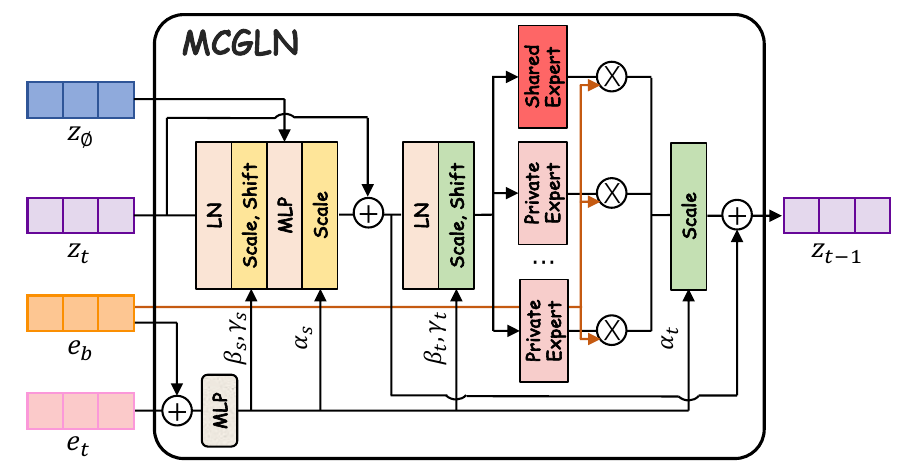}
\vspace{-0.3cm}
\caption{The architecture of MCGLN. The behavior-agnostic preference $z_{\varnothing}$ serves as the transfer subject. The target behavior $b_t$ guides the direction as hard router for MoE. Condition representation $e_b$ and $e_t$ generate adaptive weights.}
\vspace{-0.5cm}
\label{fig:mcgln}
\end{center}
\end{figure}

The latent diffusion model aims to achieve the transfer of preference presentation from behavior-agnostic to behavior-specific, which allows for the full utilization of prior knowledge of the diverse behavior information to capture user intentions precisely~\cite{dimerec}. The pre-trained MBAE in has already sufficiently defined the latent space encompassing possible user preferences, then our LDM performs both forward and reverse process within this latent space.

Given the user's multi-behavior interaction sequence, we can obtain the contextual encoding representation $z_{\varnothing}$ and $z_b$ at the item masked position by masking or preserving the behavior types, which serves as the start and target of our transfer task. 

We use DDIM~\cite{ddim} as our base diffusion algorithm, where the forward process adds varying degrees of noise to the target representation to obtain a series of intermediate noise states to match uncertainty modeling~\cite{diffurec}. Based on Eq.\ref{forward_process} and reparameter trick, we could generate $x_t^b$ directly as follows:
\begin{equation}
\label{ldm_forward}
    z_t^b = \sqrt{\bar{\alpha_t}}z_0^b + \sqrt{1-\bar{\alpha_t}} \epsilon,
\end{equation}
where $z_0^b = z_b$ is the original representation.

Then the reverse process, as the core phase of diffusion model, aims at progressively eliminating the noise from intermediate noise states to reconstruct the original preference representation. Notably, to achieve the transition from behavior-agnostic representation to specific behavior-specific representation, the essential preference $z_{\varnothing}$ and the target behavior to be predicted $b$ naturally serve as the condition in the reverse process, with the guided denoise module as $\epsilon_\theta(z_t, t, z_{\varnothing}, b)$ and objective of Eq.\ref{objective_ldm} rewritten as:
\begin{equation}
\label{objective_new_ldm}
    L_{LDM} := \mathbb{E}_{\mathcal{E}(x), \epsilon \sim \mathcal{N}(0,\mathbf{I}), t, \mathcal{B}} \left[ \| \epsilon - \epsilon_\theta(z_t, t, z_{\varnothing}, b) \|^2 \right],
\end{equation}
which is optimized through Mean-Square Error (MSE) loss.The details of network design can be fund in Section.\ref{sec:mgcln}.

Inspired by classifier-free guidance diffusion~\cite{classifier-free} can better balance diversity, we similarly employ this strategy to realize coordinated behavior guidance. That is, during the training stage, behavior conditions are set to null $\varnothing$ with a certain probability $p = 0.2$~\cite{classifier-free}, while during inference, conditional and unconditional generation are balanced with the certain factor $\omega$:
\begin{equation}
    \epsilon_\theta^* = (1+\omega)\epsilon_\theta(z_t, t, z_{\varnothing}, b) - \omega\epsilon_\theta(z_t, t, z_{\varnothing}, \varnothing).
\end{equation}
Note that the behavior-agnostic preference serving as the transfer subject is consistently maintained. Thus, our model achieves flexible generation of target behavior-specific preference representations, which then be decoded by MBAE to obtain predicted items.

\subsection{Multi-Condition Guided Layer Norm} \label{sec:mgcln}

Previous conditional denoising methods~\cite{dreamrec, dmcdr} amalgamate and compress guiding conditions together, combining with noised information through networks to achieve the guidance generation. However, these approaches uniformly processes multiple constraint information, failing to adequately preserve the semantic characteristics in multi-conditional scenarios. 
Therefore, we propose \textbf{Multi-Condition Guided Layer Normalization (MCGLN)} block as the denoiser $\epsilon_\theta$, which follows the design of adaptive layer norm (AdaLN)~\cite{adaln, dit} for stable denoising, and combines with Mixture-of-Experts (MoE)~\cite{moe, mmoe} technique for behavior selection. Furthermore, we weight the importance of conditional information and plan separate guidance modulation~\cite{dcrec}. The architecture is depicted in Figure \ref{fig:mcgln}.

Specifically, since the behavior-agnostic preference representation $z_{\varnothing}$ serves as the transfer subject, it naturally acts as the primary condition. Therefore, we directly concatenate it with the noisy latent representation $z_t$ and map the combined vector through an MLP, which ensures the primary influence permeates the entire denoising process~\cite{mcm-ldm}.

Behavior information is utilized to guide the generation toward the target behavior $b_t$ direction. Therefore, we rely on Customized Gate Control (CGC)~\cite{ple}, which incorporates both shared and private experts based MoE and can effectively disentangle the conflicts among distinct behaviors. Unlike existing approaches ~\cite{dit-moe, ec-dit} that introduce MoE into diffusion models, which utilizes soft routing with sparsity characteristic aiming at scaling, we focus on disentangling the multi-task scenario. Hence we treat the target behavior type as a hard router to determine the selection of private experts, then combining with shared experts to obtain the final output:
\begin{equation}
    w(x) = \text{softmax}(W_gx),
\end{equation}
\begin{equation}
    \text{MoE}(x, b_t) = w(x)\text{Stack}([e_{s,1},\dots,e_{s,m_s}, e^{b_t}_{p,1},\dots,e^{b_t}_{p,m_p}]),
\end{equation}
where $W_g \in \mathbb{R}^{({m_s+m_p})\times d}$ is the gate weight matrix, $m_s$ and $m_p$ are the numbers of shared experts $e_{s}$ and private experts $e_{p}^{b_t}$ respectively, where $b_t$ indicates the corresponding behavior.

In addition, time step provides progress information for the denoising to guide phased strategy adjustments~\cite{ddpm}. Therefore, we align it $e_t$ to the same dimension as the behavior representation $e_{b_t}$ and combine to produce the scale and shift parameters $\alpha$, $\beta$ and $\gamma$ for the residual adaptive normalization layers~\cite{disrt}. Note that $b_t=\varnothing$ may occur due to classifier-free guidance, thus in such case, the private experts are discarded and directed to the shared experts:
\begin{equation}
    \text{MoE}(x, \varnothing) = w(x)\text{Stack}([e_{s,1},\dots,e_{s,m_s}]).
\end{equation}

The overall module process can be regarded as:
\begin{equation}
    \hat{z_t} = z_t + \alpha_s\text{MLP}(\text{LN}(\text{Concat}(z_t+z_{\varnothing}))\gamma_s + \beta_s),
\end{equation}
\begin{equation}
    z_{t-1} = \hat{z}_t + \alpha_t\text{MoE}(\text{LN}(\hat{z}_t)\gamma_t + \beta_t, b_t),
\end{equation}
\begin{equation}
    where \quad \alpha_t, \beta_t, \gamma_t, \alpha_s, \beta_s, \gamma_s = \text{MLP}(e_t + e_{b_t}).
\end{equation}
Through MCGLN, the denoiser $\epsilon_\theta$ achieves the combination of multi-condition information across different dimensions, which is crucial for complex contextual perception.

\subsection{Training and Inference} \label{section_training}

\textbf{Training.} We divide the model training into three stages. First, we conduct pre-training for the encoder and decoder of MBAE using Cloze task under the loss of Eq.\ref{objective_mbae}, with the propose of this stage primarily in constructing the unified latent space of user preferences. Second, we freeze the pre-trained encoder and decoder then train the diffusion model in the latent space aiming at the transition from behavior-agnostic preferences to target behavior-specific preferences, thereby obtaining a grasp of user preferences that incorporates broader prior knowledge. In this process, our optimization objective is the reconstruction loss in Eq.\ref{objective_ldm}.In the third stage, we freeze the latent diffusion model while unfreezing the decoder, and switch from the Cloze task to the next item prediction task, to enable the perception of sequential relationship~\cite{eidp} through fine-tuning. This adaptive learning can also effectively bridge the bias between the diffusion model and the decoder, for the precise spatial transfer after divided training.

\noindent \textbf{Inference.} The inference process aims to generate the next possible item given target behavior based on the user's multi-behavior interaction sequence. Therefore, we start from random sampling $z_T$ in Gaussian noise, utilizing the user's behavior-agnostic preference representation $z_{\varnothing}$ obtained by the encoder, and gradually denoise for $T$ steps through the MCGLN denoiser to produce the user's behavior-specific representation $z_b = z_0$ under the target behavior $b$, which is then transferred from latent preference space into item space through the decoder. Consistent with~\cite{dreamrec}, we implement the generation of item representation, which enables entry into the entire item space for similarity search to produce the recommendation list, thereby transcending the confines of candidate sets.

The pseudo code of training and inference phase are both demonstrated in Appendix.\ref{app:pseudo_code}.

\section{Experiments}

In this section, we conduct extensive experiments on several public datasets to answer the following questions:
\begin{itemize}
    \item \textbf{RQ1:} How does our model perform in multi-behavior sequential recommendation scenario compared with the existing methods?
    \item \textbf{RQ2:} How do different components of our model contribute to the final performance?
    \item \textbf{RQ3:} How does our model leverage the latent space to capture the precise preference ?
    \item \textbf{RQ4:} How does the performance of our model vary with different hyperparameter settings?
\end{itemize}

\subsection{Experimental Settings}

\subsubsection{\textbf{Dataset.}} 

We evaluate the models based on three widely used datasets, which have already been considered standard benchmarks in MBSR scenario, with the brief introduction as following: 
\begin{itemize}
    \item \textbf{Yelp:} This dataset is collected from Yelp challenge. Based on explicit ratings, user interactions are categorized into four types, \textit{i.e.}, \textit{dislike}, \textit{neutral}, \textit{like} and \textit{tip}.
    \item \textbf{Retail:} This dataset is collected from Taobao, one of the world's largest e-commerce platforms. There are four types of behaviors, \textit{i.e.}, \textit{click}, \textit{add-to-favorite}, \textit{add-to-cart} and \textit{purchase}. Some baselines also refer to this dataset as Taobao/Tmall.
    \item \textbf{IJCAI:} This dataset is released by IJCAI competition for user activity modeling from an online e-commerce platform, It also includes four types of behaviors, \textit{i.e.}, \textit{click}, \textit{add-to-favorite}, \textit{add-to-cart} and \textit{purchase}.
\end{itemize}

For fair comparison, we adopt a similar preprocessing to previous baselines~\cite{mb-gmn, mb-str}, where each user sequence containing the last interactive serves as the test dataset, with the rest for training. Note the difference is that we don't need to explicitly specify a fixed target behavior (\textit{e.g.}, \textit{like} or \textit{purchase} in baselines), and thus the test set contains multiple possible behaviors. Besides, we don't need to construct candidate sets based on negative sampling. The statistics of the datasets can be seen in Table \ref{tab:dataset}.

\begin{table}[t!]
\setlength{\abovecaptionskip}{2pt}
\vspace{-0.3cm}
\caption{The statistics of  datasets}
\label{tab:dataset}
\centering
\begin{small}
\setlength\tabcolsep{2.5pt} 
  \begin{tabular}{ccccc}
    \toprule
    \textbf{Dataset} & \textbf{User} & \textbf{Item} & \textbf{Interactions} & \textbf{Behavior Type} \\
    \midrule
    Yelp & 19,800 & 22,734 & $1.4 \times 10^6$ & Tip, Dislike, Neutral, Like \\
    Retail & 147,894 & 99,037 & $7.6 \times 10^6$ & Click, Favorite, Cart, Purchase \\
    IJCAI & 423,423 & 874,328 & $3.6 \times 10^7$ & Click, Favorite, Cart, Purchase \\
    \bottomrule 
  \end{tabular}
\end{small}
\vspace{-0.3cm}
\end{table}

\subsubsection{\textbf{Evaluation Metrics.}}
We follow top-K protocol and adopt two widely used ranking metrics, \textit{Recall@K} and \textit{NDCG@K}~\cite{ndcg} to evaluate the performance of our model. The higher the values of the both metrics, the better the model perform.

\begin{table*}[t!]
  \caption{The overall performance comparison of different models on three datasets. The best results are \textbf{boldfaced} and the second-best results are \underline{underlined}, with the relative improvements denoted as Improve$\uparrow$. All baseline results are derived from reproduced experiments referencing the original papaers and released code.}
  \vspace{-0.3cm}
  \setlength{\tabcolsep}{3.5pt}
  \label{tab:overall}
  \begin{tabular}{@{}lcccccccccccc@{}}
    \toprule
    \multirow{2.5}{*}{\textbf{Models}} & \multicolumn{4}{c}{\textbf{Yelp}}  & \multicolumn{4}{c}{\textbf{Retail}} & \multicolumn{4}{c}{\textbf{IJCAI}}
    \\ \cmidrule(r){2-5} \cmidrule(r){6-9} \cmidrule(r){10-13} & R@10 & N@10 & R@20 & N@20 & R@10 & N@10 & R@20 & N@20 & R@10 & N@10 & R@20 & N@20\\
    \midrule
    BERT4Rec & 0.0501 & 0.0245 & 0.0828 & 0.0331 & 0.1414 & 0.0811 & 0.1853 & 0.0921 & 0.2386 & 0.1473 & 0.3009 & 0.1632 \\
    DiffuRec & 0.0526 & 0.0280 & 0.0840 & 0.0362 & 0.1476 & 0.0860 & 0.1914 & 0.0971 & 0.2811 & 0.1702 & 0.3405 & 0.1816 \\
    \midrule
    MBGCN & 0.0497 & 0.0229 & 0.0796 & 0.0323 & 0.1398 & 0.0801 & 0.1832 & 0.0910 & 0.2274 & 0.1401 & 0.2937 & 0.1590 \\
    MB-GMN & 0.0517 & 0.0257 & 0.0859 & 0.0343 & 0.1434 & 0.0829 & 0.1859 & 0.0914 & 0.2408 & 0.1488 & 0.3072 & 0.1654 \\
    \midrule
    MB-STR & \underline{0.0545} & 0.0275 & \underline{0.0938} & 0.0372 & 0.1532 & 0.0887 & 0.1990 & 0.1002 & \underline{0.2904} & \underline{0.1999} & \underline{0.3430} & \underline{0.2132} \\
    PBAT & 0.0432 & 0.0207 & 0.0771 & 0.0291 & 0.1406 & 0.0820 & 0.1821 & 0.0924 & 0.2612 & 0.1625 & 0.3223 & 0.1780 \\
    MISSL  & 0.0541 & \underline{0.0283} & 0.0922 & \underline{0.0376} & 0.1485 & 0.0854 & 0.1946 & 0.0970 & 0.2786 & 0.1792 & 0.3380 & 0.1942 \\
    M-GPT & 0.0528 & 0.0250 & 0.0915 & 0.0346 & \underline{0.1593} & \underline{0.0908} & \underline{0.2014} & \underline{0.1035} & 0.2847 & 0.1826 & 0.3388 & 0.1956 \\
    \midrule
    \textbf{FatsMB} & \textbf{0.0657} & \textbf{0.0360} & \textbf{0.1034} & \textbf{0.0454} & \textbf{0.1860} & \textbf{0.1103} & \textbf{0.2342} & \textbf{0.1223} & \textbf{0.3305} & \textbf{0.2160} & \textbf{0.3981} & \textbf{0.2331} \\
    \textbf{Improve $\uparrow$} & \textbf{+20.56\%} & \textbf{+27.21\%} & \textbf{+10.23\%} & \textbf{+20.74\%} & \textbf{+16.76\%} & \textbf{+21.48\%} & \textbf{+16.29\%} & \textbf{+18.16\%} & \textbf{+13.81\%} & \textbf{+8.05\%} & \textbf{+16.06\%} & \textbf{+9.33\%} \\
    \bottomrule
  \end{tabular}
  \vspace{-0.1cm}
\end{table*}

\subsubsection{\textbf{Baselines.}}

We compare our model with the state-of-the-art models from three lines of research topics: \textbf{Single-Behavior Sequential Models} include: BERT4Rec~\cite{bert4rec}, DiffuRec~\cite{diffurec}. \textbf{Multi-Behavior Non-Sequential Models} include: MBGCN~\cite{mbgcn}, MB-GMN~\cite{mb-gmn}. \textbf{Multi-Behavior Sequential Models} include: MB-STR~\cite{mb-str}, PBAT~\cite{pbat}, MISSL~\cite{missl}, M-GPT~\cite{m-gpt}. Details of baselines refer to Appendix.\ref{app:baseline}.

\subsubsection{\textbf{Parameters Settings.}}

Our model is implemented in PyTorch~\cite{pytorch}, and optimized by the AdamW~\cite{adamw} optimizer with learning rate selected as $2e^{-3}$. We set the batch size to $256$ for \textbf{Yelp} and \textbf{Retail}, while $64$ for \textbf{IJCAI}. Also, we set the embedding dimension to $d=64$, the max sequence length to $L=50$ and dropout rate to $0.1$. There are 2 Transformer layers in encoder, as well as 2 MCGLN layers in denoiser. The study of the effect of hyperparameters taking values from the range will be visible later. All experiments are conducted on the 32G NVIDIA V100 GPUs.

\subsection{Overall Performance (RQ1)}

To show the effectiveness of our model, we conduct experiments on three public datasets, with detailed results shown in Table ~\ref{tab:overall}. We summarize the following observations:
\begin{itemize}
    \item FatsMB generally outperforms all the baselines on the three datasets, with the improvement up to 20.56\% of \textit{Recall@10} and 27.21\% of \textit{NDCG@10}. We ascribe the performance gains of our model to the following points: i) The autoencoder of FatsMB is based on the Transformer backbone, with a novel joint integration strategy of behavior and position information incorporated during the attention computation phase, which effectively enhances the model's capability to capture sequential correlations; ii) The construction of the unified preference latent space, within the transfer of preference from behavior-agnostic to behavior-specific, enables the generation of target preference to incorporate extensive prior knowledge; iii) The ability of generative paradigm to directly generate target items helps achieve optimal recommendations in large item scale without candidate sets.
    \item Whether the information of multi-behavior or sequence can effectively improve model performance. This is for the former enables information classification and attribution coupling, while the latter allows learning temporal evolution and transitions. Both operations make the model more suitable for MBSR scenario.
    \item The introduction of diffusion mode (\textit{e.g.}, DiffuRec) also brings improvements in metrics, demonstrating the effectiveness of diffusion models in the recommendation domain. 
    Because this generative approach naturally supports exploring new recommendation possibilities without being constrained by historical fitting.
    Therefore, our leverage of latent diffusion model in the MBSR scenario, along with unique design to adapt to domain task, represents a beneficial and valuable attempt.
\end{itemize}

\subsection{Components Contributions (RQ2)}

\begin{table}[t!]
  \caption{The performance comparison of ablation for different modules with different components. Comparison can be divided into inter-module and intra-module. The best results of diverse module variants are boldfaced.}
  \vspace{-0.3cm}
  \setlength{\tabcolsep}{1.5pt}
  \label{tab:ablation}
  \begin{tabular}{c@{\hspace{0.5mm}}l@{\hspace{0.5mm}}cccccc}
    \toprule
    \multicolumn{2}{c}{\multirow{2.5}{*}{\textbf{Variants}}} & \multicolumn{2}{c}{\textbf{Yelp}} & \multicolumn{2}{c}{\textbf{Retail}} & \multicolumn{2}{c}{\textbf{IJCAI}} \\
    \cmidrule(r){3-4} \cmidrule(r){5-6} \cmidrule(r){7-8} & & R@20 & N@20 & R@20 & N@20 & R@20 & N@20 \\
    \midrule
    \multirow{3}{*}{MBAE} & \multicolumn{1}{|l}{\textit{w} APE}& 0.0936 & 0.0381 & 0.2183 & 0.1103 & 0.3701 & 0.2105 \\
    & \multicolumn{1}{|l}{\textit{w} RoPE} & 0.0981 & 0.0430 & 0.2252 & 0.1170 & 0.3712 & 0.2096 \\
    & \multicolumn{1}{|l}{\textit{w} BaRoPE} & \textbf{0.1003} & \textbf{0.0438} & \textbf{0.2280} & \textbf{0.1189} & \textbf{0.3880} & \textbf{0.2227} \\
    \midrule
    \multirow{3}{*}{FatsMB} & \multicolumn{1}{|l}{\textit{w} MLP} & 0.0980 & 0.0432 & 0.2279 & 0.1197 & 0.3798 & 0.2211 \\
    & \multicolumn{1}{|l}{\textit{w} AdaLN} & 0.1019 & 0.0447 & 0.2286 & 0.1195 & 0.3857 & 0.2253 \\
    & \multicolumn{1}{|l}{\textit{w} MCGLN} & \textbf{0.1034} & \textbf{0.0454} & \textbf{0.2342} & \textbf{0.1223} & \textbf{0.3981} & \textbf{0.2331} \\
    \bottomrule
  \end{tabular}
  \vspace{-0.3cm}
\end{table}

To verify the effectiveness of each module and component in FatsMB and their advantages over existing methods, we conduct extensive module ablation and comparative experiments, with results shown in Table \ref{tab:ablation}. We analyze from the following three perspectives.

\subsubsection{\textbf{Effect of Latent Space.}} For the effectiveness of transfer learning in the latent space ,we conduct ablation experiments on the latent diffusion model, comparing the pre-trained MBAE with the complete model FatsMB. The results are shown in inter-module comparison of Table \ref{tab:ablation}, which demonstrated the further operations in the latent space indeed 
achieve precise user preference comprehension with extensive prior knowledge and accurate guidance.

\subsubsection{\textbf{Effect of BaRope.}} For sequential information integration and relevance capture among item, behavior and position, we compared commonly used position embedding methods by designing three variants: MBAE with Absolute Position Embedding (APE), Rotary Position Embedding (RoPE) and  our proposed Behavior-aware RoPE (BaRoPE). The results are shown in MBAE section of Table \ref{tab:ablation}, through which the effective performance improvement brought by BaRoPE can be observed. Therefore, we can conclude that the introduction of relative position relationships can capture key information in history interaction sequences, while the combination of behavior information with relative position can also precisely grasp user intentions under various behavior contexts.

\begin{figure}[t!]
    \vspace{-0.2cm}
    \centering
    \subfloat[Matrix of APE \label{fig:ape}]{
        \includegraphics[width=4cm]{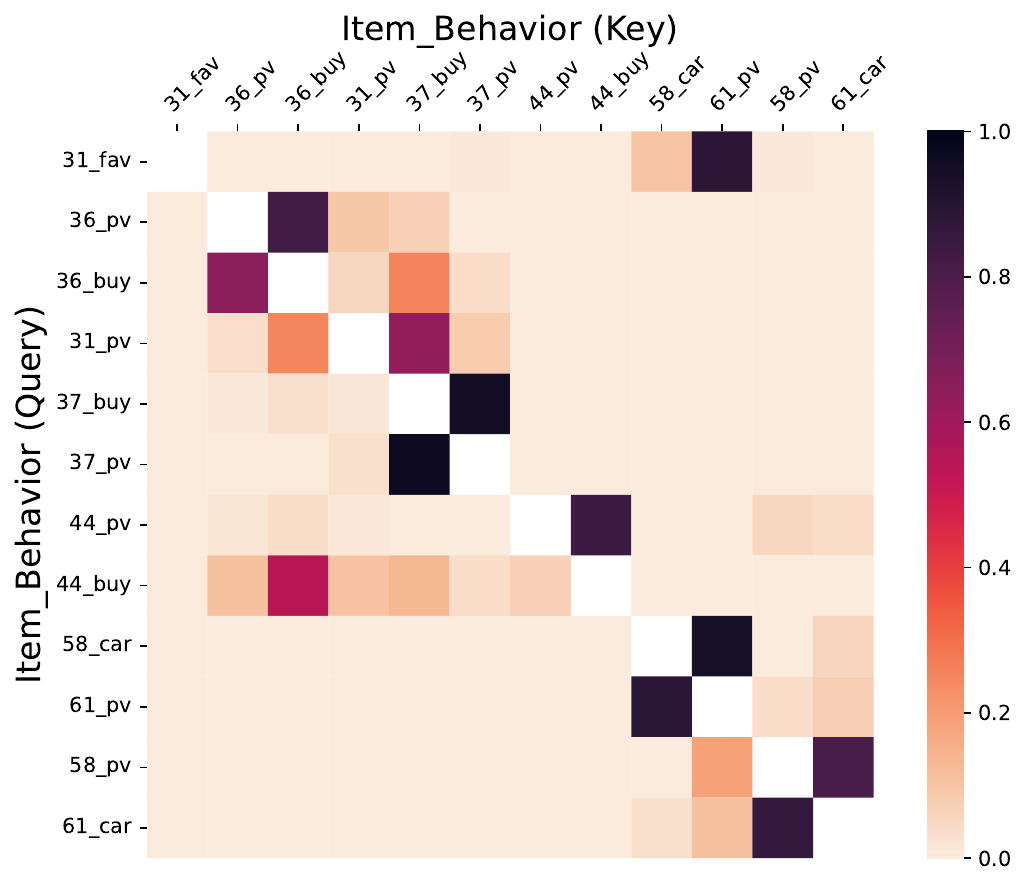}
    }
    \subfloat[Matrix of BaRoPE \label{fig:barope}]{
        \includegraphics[width=4cm]{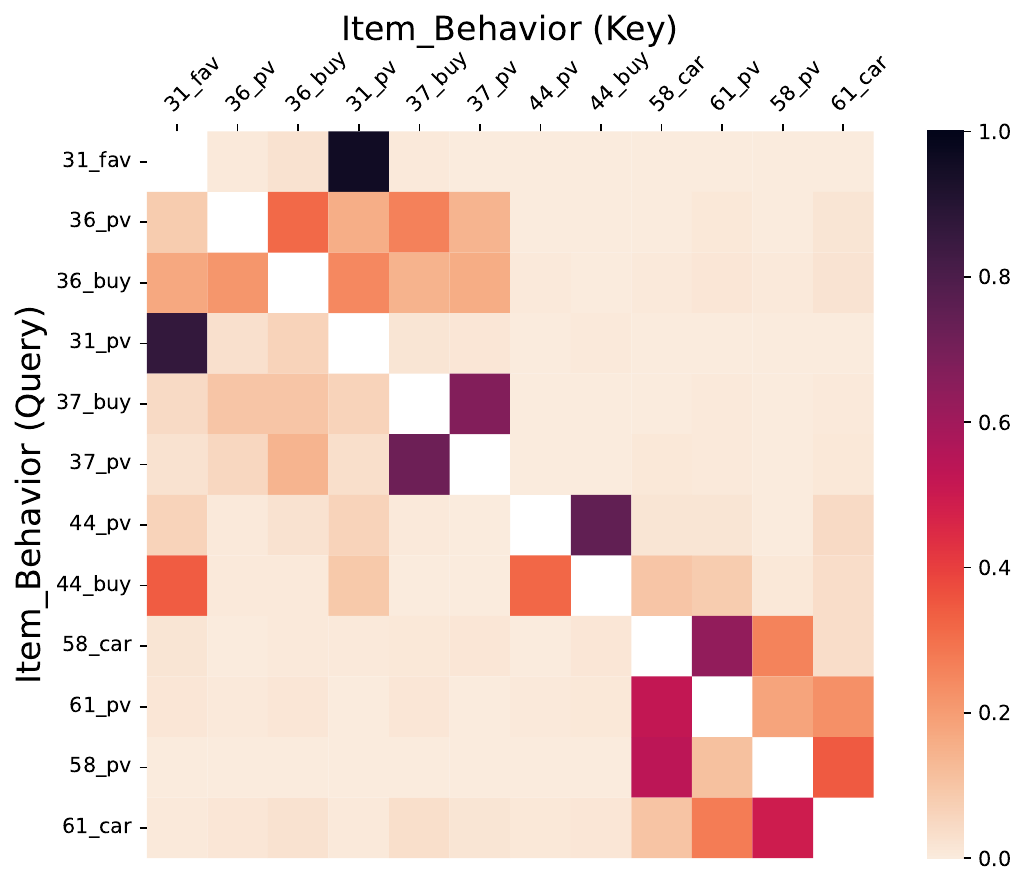}
    }
    \vspace{-0.3cm}
    \caption{The comparison of attention matrics of APE and BaRoPE on \textit{user\_1} in \textbf{Retail} dataset. The sequential arrangement of \textbf{Item\_Behavior} means the user interacted with \textit{item} through \textit{behavior}.}
    \vspace{-0.3cm}
    \label{fig:attention}
\end{figure}

Furthermore, we conduct visualization of attention distributions based on the interaction sequence of \textit{user\_1} in the \textbf{Retail} dataset, intuitively presented in Figure \ref{fig:attention}. Specifically, we employed the average of multi-layer multi-head attention scores for visualization. The observations show that compared to APE, BaRoPE has two distinct advantages: i) a larger correlation region and superior symmetry that enable more effective and comprehensive integration of both short- and long-term information across the sequence; ii) more precise focus on the successive behaviors between identical items (\textit{31\_fav} and \textit{31\_pv}, 5\textit{8\_car} and \textit{58\_pv}, \textit{etc}), effectively combining and capturing the casual patterns between behaviors and items.

\subsubsection{\textbf{Effect of MGCLN.}} For the architecture selection of multi-condition guided denoising module, we explore three commonly used variants: Multilayer Perceptron (MLP), Adaptive Layer Norm (AdaLN) and our proposed Multi-condition Guided Layer Norm (MCGLN), with result shown in FatsMB section of Table \ref{tab:ablation}. It can be observed that concatenating and merging multiple condition information followed by MLP performs worst, sometimes even failing to achieve the performance of the pre-trained model. AdaLN obtains certain improvements due to the application of adaptive normalization. Our MCGLN, through separate processing and adaptive balancing of multiple condition information and further adaptation to multi-tasks via MoE, achieves the best performance.

\subsection{Effect Study (RQ3)}

\subsubsection{\textbf{Preference Distribution Experiment.}} 
\begin{figure}[t!]
    \vspace{-0.3cm}
    \centering
    \hspace{0.2cm}
    \subfloat[Distribution Visualization \label{fig:multi_pre}]{
        \includegraphics[width=3.5cm]{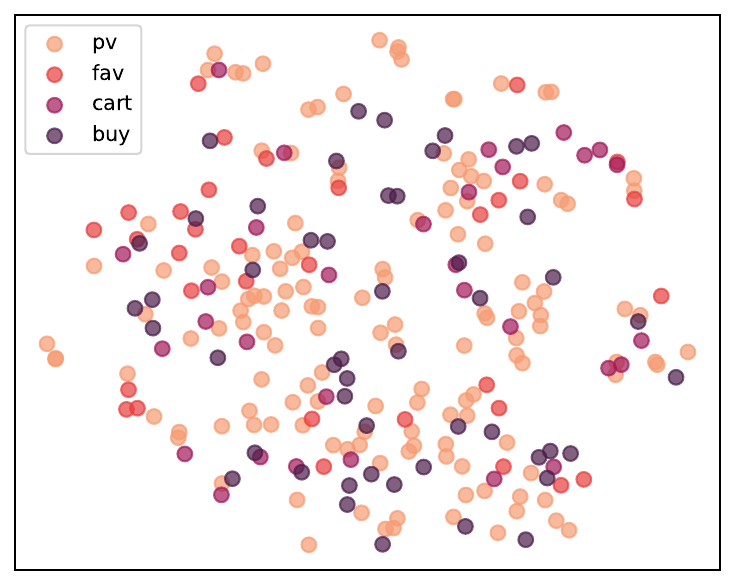}
    }
    \subfloat[Preference Similarity \label{fig:sim_pre}]{
        \includegraphics[width=4.3cm]{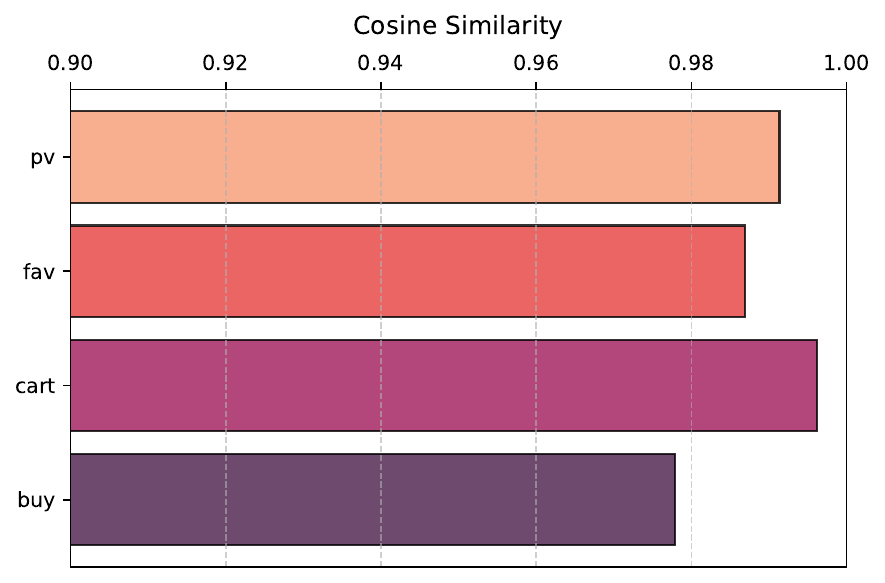}
    }
    \vspace{-0.3cm}
    \caption{The preference distribution and similarity experiment in \textbf{Retail} dataset.}
    \vspace{-0.3cm}
    \label{fig:preference_distribution}
\end{figure}

To demonstrate the effective of unified multi-behavior preference modeling, we select a batch of user interaction sequences from \textbf{Retail} dataset and encode into behavior-specific preferences in the latent space through our model. Then the representations are visualized via t-SNE~\cite{tsne} dimensionality reduction, as shown in Figure \ref{fig:multi_pre}. It can be observed that the preference representations specific to different behaviors are uniformly distributed within a shared latent space, thereby proving the effective unified embedding of our model.

However, a potential concern is that random representations might also achieve such uniform distribution. Thus we further compute the cosine similarity between each user's behavior-agnostic and various behavior-specific preferences. The high correlation shown in Figure \ref{fig:sim_pre} validate the fidelity of information capture and simultaneously suggests the feasibility of preference transfer.

\begin{figure}[t!]
    \vspace{-0.2cm}
    \centering
    \subfloat[Comparison of Recall \label{fig:fewshot_recall}]{
        \includegraphics[width=4cm]{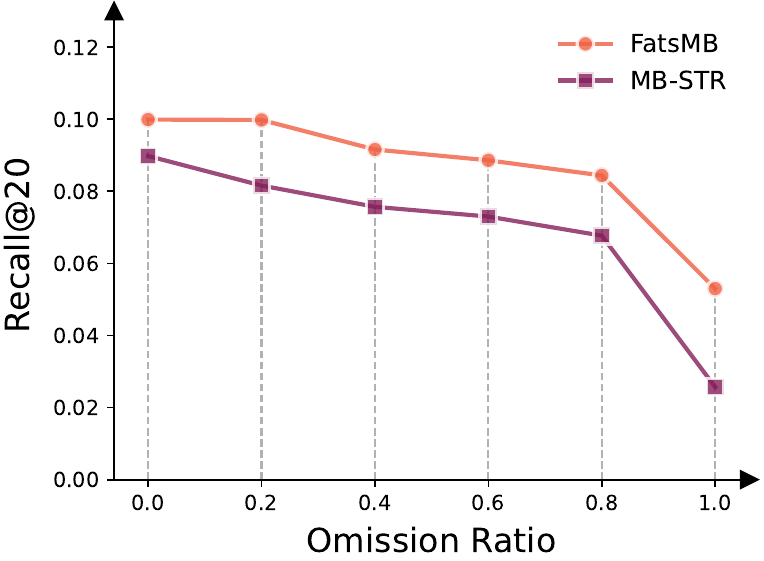}
    }
    \subfloat[Comparison of NDCG \label{fig:fewshot_ndcg}]{
        \includegraphics[width=4cm]{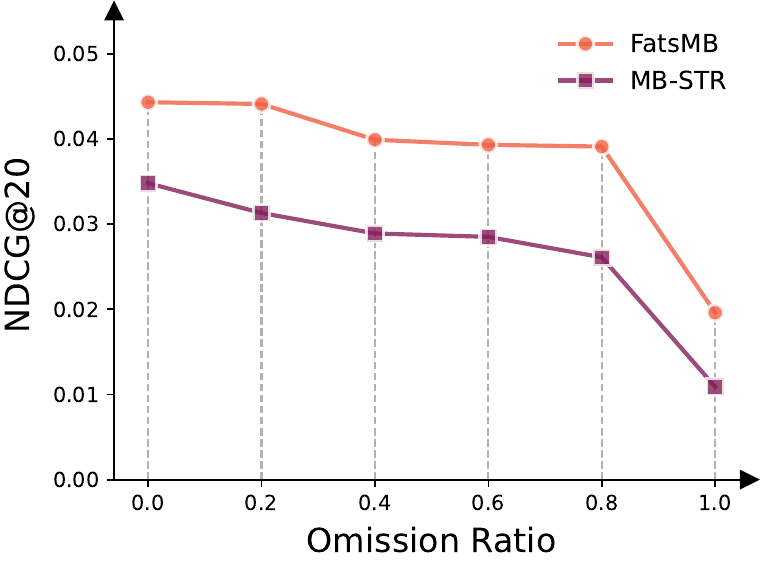}
    }
    \vspace{-0.1cm}
    \caption{The comparison of FatsMB and MB-STR in the few-shot experiment in \textbf{Yelp} dataset.}
    \vspace{-0.5cm}
    \label{fig:fewshot}
\end{figure}

\subsubsection{\textbf{Few-shot Experiment.}} To evaluate our model's ability to leverage unified representations in latent space for sparse behavior prediction, we conduct few-shot experiments to prove the knowledge transfer between behaviors. Specifically, we treat \textit{like} in \textbf{Yelp} dataset as the target behavior, and randomly drop its interactions by an omission ratio during training (up to $1$ as zero-shot scenario), while maintaining the prediction dataset.

The result is presented in Figure \ref{fig:fewshot}, with consistent trends observed across other datasets. We draw conclusions from three aspects: i) Wether for behavior-specific or behavior-fixed prediction, our model performs better than baselines; ii) Mild data omission (\textit{i.e.}, $0.2$) affects our model little, but obviously impacts on baselines; and our model exhibits a relatively smaller decay rate overall; iii) Even in zero-shot scenario, our model retains certain capability, in contrast to the near-collapse of baseline. We attribute all these to the effective fusion and transfer of unified latent space, compensating for the lack of target behavior with external information.

\begin{figure*}[ht!]
    \centering
    \subfloat[Item Mask Probability $\rho$ \label{fig:item}]{
        \includegraphics[width=3.6cm]{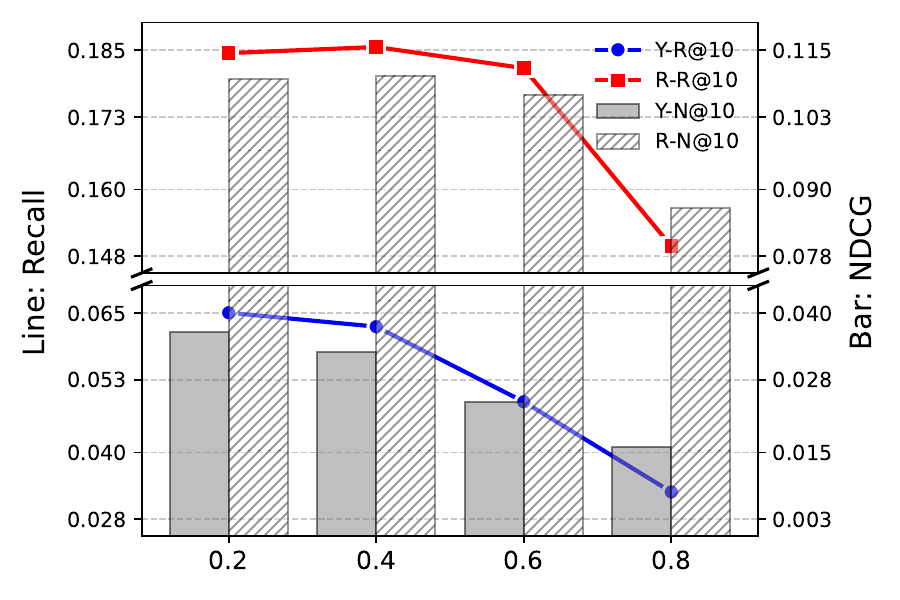}
    }
    \hspace{-0.34cm}
    \subfloat[Behavior Mask Probability $\sigma$ \label{fig:behavior}]{
        \includegraphics[width=3.6cm]{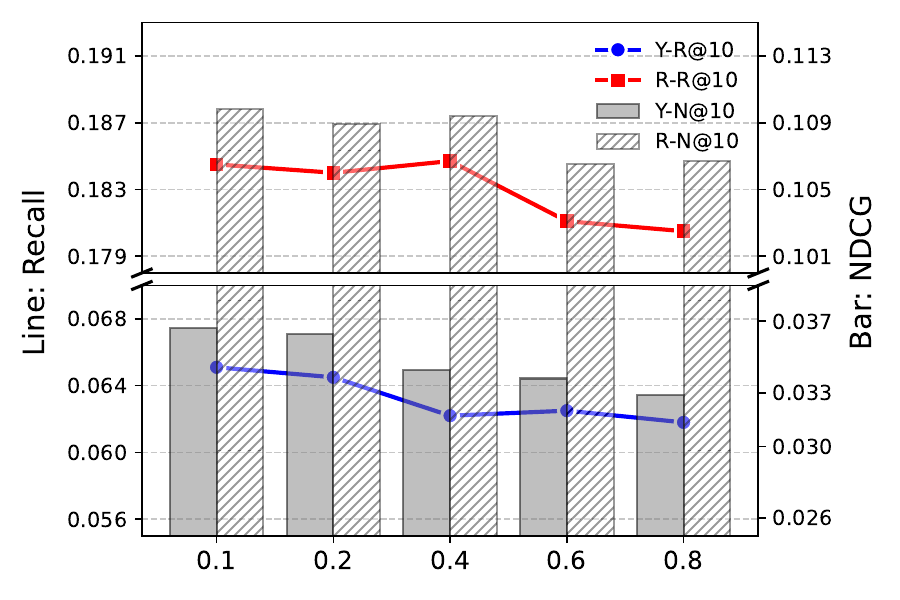}
    }
    \hspace{-0.34cm}
    \subfloat[Diffusion Steps $T$ \label{fig:step}]{
        \includegraphics[width=3.6cm]{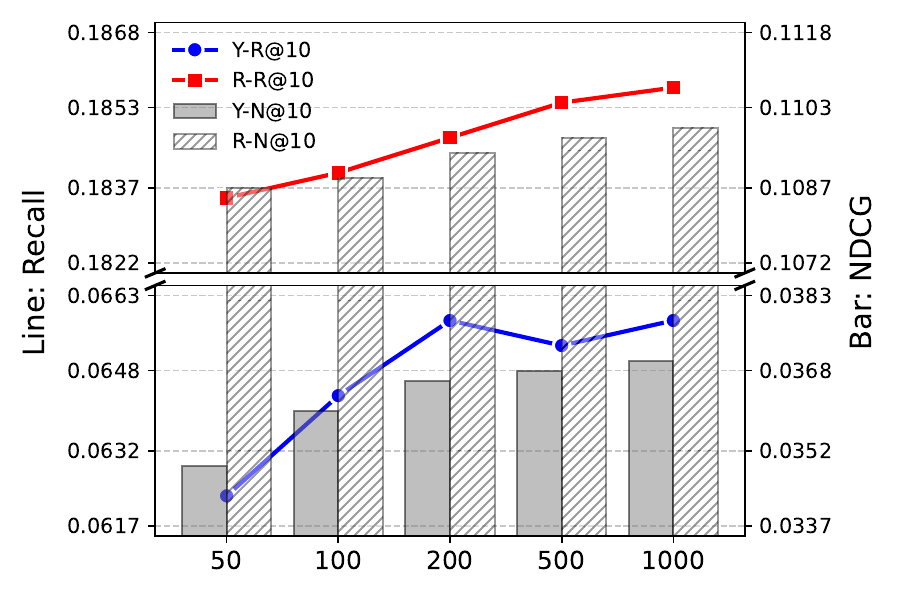}
    }
    \hspace{-0.34cm}
    \subfloat[Time Step Interval $\Delta t$ \label{fig:jump}]{
        \includegraphics[width=3.6cm]{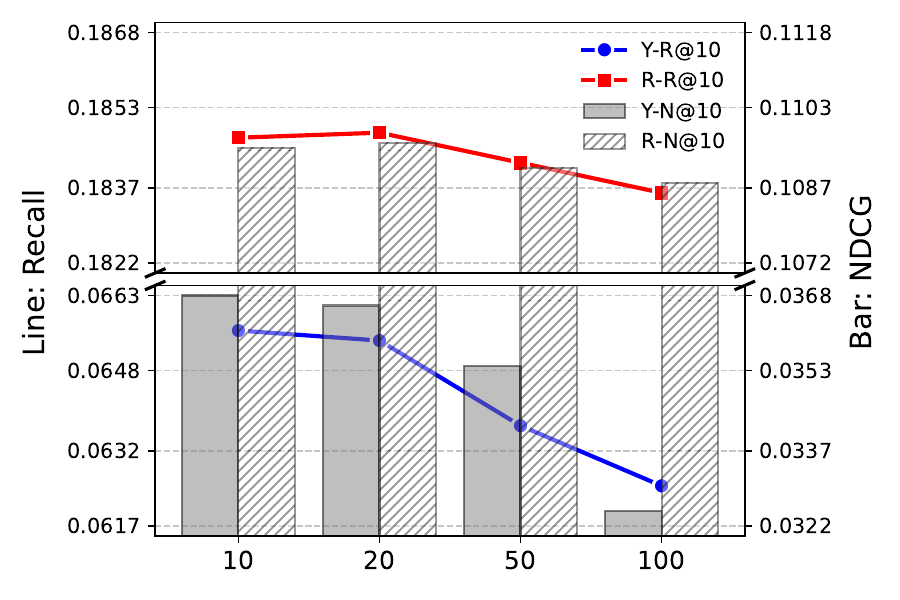}
    }
    \hspace{-0.34cm}
    \subfloat[Guidance Strength $\omega$ \label{fig:free}]{
        \includegraphics[width=3.6cm]{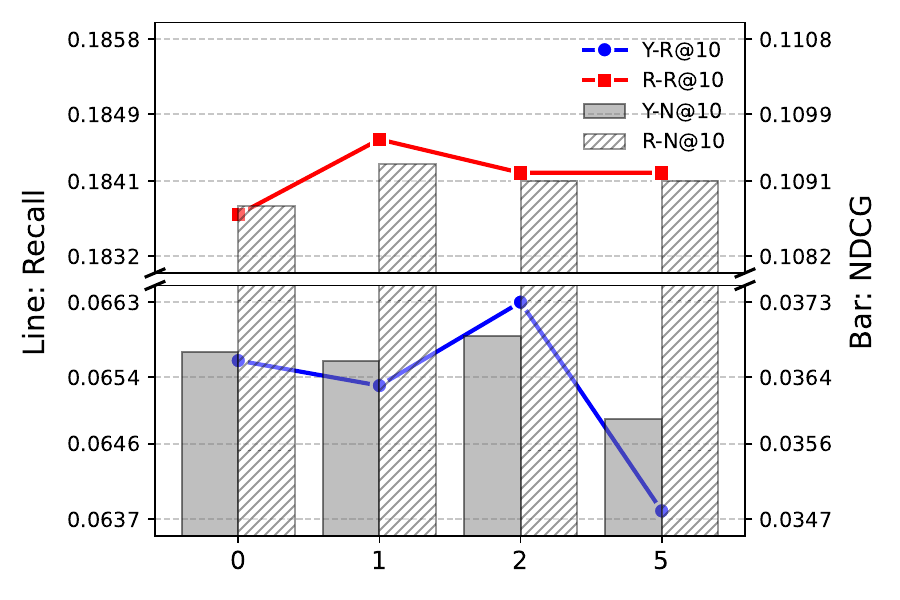}
    }
    \vspace{-0.3cm}
    \caption{The performance comparison of different hyperparameters on Yelp and Retail datasets. 
    \textit{Y-R} and \textit{R-R} indicates \textit{Recall} on \textbf{Yelp} and \textbf{Retail} datasets. \textit{Y-N} and \textit{R-N} indicate \textit{NDGC} on \textbf{Yelp} and \textbf{Retail} datasets. 
    Among these, (a)(b) are the results on MBAE and (c)(d)(e) are the results on FatsMB.
    }
    \vspace{-0.3cm}
    \label{fig:hyper}
\end{figure*}

\subsubsection{\textbf{Time and Space Complexity Analysis.}}

During training, the time complexity of our model is dominated by the multi-head attention of MBAE $O(L^2d+Ld^2)$, and MCGLN of LDM $O(Ld^2)$. During inference, the LDM requires multiple denoising steps, resulting in an overall time complexity of $O(L^2d+(1+T/\Delta t)L d^2)$.
As for space complexity, the learnable parameters of our model mainly focus on item and behavior embeddings $O((\lvert \mathcal{V} \rvert + \lvert \mathcal{B} \rvert)d)$, multi-head attention of MBAE $O(Ln^2 + d^2)$, and MoE of MGCLN $O((m_s + \lvert \mathcal{B} \rvert m_p)d^2)$.
These are all comparable to SOTA baselines, with comparative experiments shown in Appendix.\ref{app:complexity}.

\subsection{Hyperparameters Analysis (RQ4)}

\subsubsection{\textbf{Effect of Item Mask Probability $\rho$.}} We vary $\rho$ in the range of $[0.2,0.4,0.6,0.8]$ with the result shown in Figure \ref{fig:item}. It can be observed that a moderately small mask probability enables the model to achieve effective learning, thereby obtaining adequate pre-training performance. However, performance shows increasingly sharp degradation as $\rho$ increases, as excessive mask causes the model losing rich context information. 
This is consistent with the conclusion of previous work~\cite{bert4rec}.

\subsubsection{\textbf{Effect of Behavior Mask Probability $\sigma$.}} We evaluate the influence of $\sigma$ in the range of $[0.1,0.2,0.4,0.6,0.8]$ and present the result in Figure \ref{fig:behavior}. We can observe that the performance declines with the increase of mask probability overall. The fluctuation is not significant within small value range, during which it can be considered as a balance against overfitting. but when the value is large, due to substantial loss of effective information, specific behaviors can't be effectively learned, resulting in arresting performance degradation. Nevertheless, this remains a necessary operation for unifying multiple preferences within a single space.

\subsubsection{\textbf{Effect of Diffusion teps $T$.}} We conduct experiments on the selection of $T$ from $[50,100,200,500,1000]$, with result presented in Figure \ref{fig:step}. We can find that as diffusion step increases, the performance gradually improves, as this enables deeper and finer-grained denoising, while with a smoother noise variation. However, this consequently leads to a significant increase in inference time and a diminishing marginal effect. Considering the balance between time cost and effectiveness, we selected $200$ as the total step length in all the other experiments.

\subsubsection{\textbf{Effect of Time Step Interval $\Delta t$.}} We fix $T$ and vary $\Delta t$ in $[10,20,50,100]$. The result shown in Figure \ref{fig:jump} demonstrate that as the sampling interval increases, the model's performance deteriorates. This is reasonable, as larger jumps lead to loss of recovery details, and further accumulate increasingly errors. However, skip steps can effectively improve generation speed, thus $\Delta t = 20$ is an acceptable choice that doesn't compromise performance.

\subsubsection{\textbf{Effect of Guidance Strength $\omega$.}} The selection of $\omega$ varies in $[0,1,2,5]$, where $\omega = 0$ degrades to classifier guided strategy. Meanwhile, a higher $\omega$ can strengthen personalized guidance but may also lead to overfitting to guidance and sparse diversity generation. The result in Figure \ref{fig:free} shows that for different datasets, the optimal guidance strength may also differ, therefore it may need to be selected according to environmental changes.

\section{Related Work}

\subsection{Multi-behavior Sequential Recommendation}

Multi-behavior recommendation aims to fully utilize users' heterogeneous information to achieve item prediction under target behavior. Early methods like MBGCN~\cite{mbgcn} and MB-GMN~\cite{mb-gmn} leverage Relational Graph Neural Networks (RGNN) to implement information propagation between user and item nodes connected by interaction relationship. CML~\cite{cml} and MBSSL~\cite{mbssl} focus on the alignment of different behaviors, achieving consistency modeling by contrastive learning. However, these methods often fail to consider the temporal evolution in user preferences over time, and thus recent works pay attention to multi-behavior sequential recommendation (MBSR). MB-STR~\cite{mb-str} and M-GPT ~\cite{m-gpt} capture dependencies in multi-behavior interaction sequences from multiple perspective. PBAT~\cite{pbat} and MISSL~\cite{missl} learn personalized hierarchical multi-behavior relationships. END4Rec~\cite{end4rec} performs sequence information purification through noise-decoupling. Note that these models are all discriminative paradigms, resulting in suboptimal solutions of behavior-specific recommendation for intractability of variability and obtainment homogeneity. MBGen~\cite{mbgcn} leverages autoregressive generative model but still fail to capture the latent decision preferences. Therefore, FatsMB provides a unified solution to these issues based on LDM.

\subsection{Diffusion for Recommendation}

Diffusion Models (DM)~\cite{ddpm,improved-ddpm, ddim} are suitable for numerous generation tasks due to the powerful generation capabilities, which have achieved significant success in diverse fields like image generation~\cite{classifier-guided, classifier-free}, style transfer~\cite{mcm-ldm, styleid, stylediffusion} and text generation~\cite{diffusion-lm, planner}. Many recent works introduce the capability of DM to handle complex distributions into recommendation systems~\cite{diffusion_rec_survey, dmcdr}. As for sequential recommendation (SR), DiffuRec~\cite{diffurec}, DreamRec~\cite{dreamrec} and DimeRec~\cite{dimerec} treat target items as generation objectives while using interaction sequences as denoising guidance. DCDR~\cite{dcdr} and SeeDRec~\cite{seedrec} treat sequences as target oppositely for generating corresponding item sequences through the reverse process. DCRec~\cite{dcrec} uses contextual information as explicit and implicit conditions for dual embedding. However, there are still regrets here: on one hand, the application of DM in MBSR is still largely unexplored. MISD~\cite{misd} only applies simple diffusion for robustness without adequately adapting to the characteristics and requirements of MBSR. On the other hand, there is limited exploration of LDM in RS, but method like DiffRec~\cite{diffrec} demonstrated the effectiveness. Therefore, applying LDM for MBSR is 
worthwhile exploring.

\section{Conclusion}

In this work, we propose FatsMB, a novel approach for multi-behavior sequential recommendation based on latent diffusion model through the transfer of preferences from behavior-agnostic to behavior-specific. Specifically, we utilize a multi-behavior autoencoder with novel position encoding strategy for the construct of unified latent preference space. Subsequently, we perform the transfer of preferences in the latent space by guided diffusion model, while conducting reverse process through multi-condition guided denoiser. Finally the target behavior-specific preference is decoded to generate the recommended items. Extensive experiments on real-world datasets demonstrate the effectiveness of our model. 
In future, we will continue to explore the utilization of LDM in MBSR.

\begin{acks}
We would like to express our gratitude to the anonymous reviewers for their constructive comments. This work was supported by the National Natural Science Foundation of China (No. 62406319).
\end{acks}

\balance
\bibliographystyle{ACM-Reference-Format}
\bibliography{sample-base-extend.bib}

\appendix

\section{Pseudo Code} \label{app:pseudo_code}

\begin{algorithm}[h!]
\SetAlgoLined
\caption{The Training Phase}
\label{alg:train}
\KwIn{User Multi-behavior Sequence $S_u$}
\KwOut{Model parameters (Encoder $\phi$, Diffusion Model $\theta$, Decoder $\psi$)}
Random initialization model parameters \;
\Stage{\textbf{1:} Training the Multi-behavior Autoencoder}{
 \For{$i \leftarrow 1$ \KwTo $len(S_u)$}{
    \uIf{random $< \rho$}{
        index $\leftarrow$ i \;
        $v_i \leftarrow [\text{mask}]$ \;
        \uIf{random $< \sigma$}{
            $b_i \leftarrow [\text{mask}]$ \;
        }
    }
 }
 H $\leftarrow E_{i}$ ($S_u$) \;
 Q, K, V $\leftarrow$ BaRope($HW^Q$), BaRope($HW^K$), $HW^V$\;
 $\text{H}^l \leftarrow \text{MultiHead}(Q,K,V)$ \;
 $z_b/z_{\varnothing} \leftarrow h^l_{index}$ \;
 $v_{[\text{mask}]} \leftarrow \mathcal{D}(z_b/z_{\varnothing})$ \;
 Calculate loss: $L_{MBAE} \leftarrow \text{Eq.}\ref{objective_mbae}$ \;
}
\Stage{\textbf{2:} Training the Latent Diffusion Model}{
 Freeze $\mathcal{E}$ and $\mathcal{D}$\;
 $z_b, z_{\varnothing} \leftarrow \mathcal{E}(S_u, b), \mathcal{E}(S_u, [\text{mask}])$ \;
 Sample $t \in \text{Uniform}({1,\dots,T})$ \;
 $z^b_t \leftarrow \sqrt{\bar{\alpha_t}}z_0^b + \sqrt{1-\bar{\alpha_t}} \epsilon$ \;
 $\hat{z}_b \leftarrow  \epsilon^*_{\theta}(z^b_t, t, z_{\varnothing, b})$ \;
 Calculate loss: $L_{LDM} \leftarrow \text{Eq.}\ref{objective_new_ldm}$
}
\Stage{\textbf{3:} Fine-tuning}{
 Freeze $\mathcal{E}$ and $\mathcal{\epsilon}$, unfreeze $\mathcal{D}$ \;
 $v_t, b_t \leftarrow S_u[-1]$ \;
 $z_{\varnothing} \leftarrow \mathcal{E}(S_u[:-1], [\text{mask}])$ \;
 Infer: $v_{\text[mask]} \leftarrow \text{Alg.}\ref{alg:infer}$ \;
 Calculate loss: $L_{MBAE} \leftarrow \text{Eq.}\ref{objective_mbae}$ \;
}
Return\;
\end{algorithm}
\vspace{-0.3cm}

\begin{algorithm}[h!]
\SetAlgoLined
\caption{The Inference Phase}
\label{alg:infer}
\KwIn{User Multi-behavior Sequence $S_u$, Target Behavior $b_t$}
\KwOut{Prediction item $v$}
$z_{\varnothing} \leftarrow \mathcal{E}(S_u, [\text{mask}])$ \;
Sample $z^{b_t}_T \in \mathcal{N}(0,\mathbf{I})$ \;
\For{$t \leftarrow T$ \KwTo 1}{
 $z^{b_t}_{t-1} \leftarrow \epsilon^*_{\theta}(z^{b_t}_t, t, z_{\varnothing}, b_t)$ \;
}
$v_t \leftarrow \mathcal{D}(z^{b_t}_0)$ \;  
Return $v_t$ \;
\end{algorithm}
\vspace{-0.3cm}

\section{Calculation Details}  \label{app:entropy}

We calculate uncertainty measurement metrics from the perspective of the entire dataset. Therefore, we use the occurrence frequency of items $I$ and behaviors $B$ as the empirical probability, and the occurrence frequency of item-behavior pairs as the joint probability. Then we calculate entropy $H(\cdot)$ to measure uncertainty and mutual information $I(\cdot,\cdot)$ to measure the degree of information sharing between variables. We then further calculate conditional entropy $H(\cdot|\cdot)$ to evaluate the impact of one variable on the uncertainty of another variable.

\section{Formula Derivation}    \label{app:derivation}

We know that RoPE borrows the definition of multiplication in complex numbers domain for the expansion of Eq.\ref{eq:rope}:
\begin{equation}
    f(q,m) = R_f(q,m)e^{i\Theta_f(q,m)} = qe^{im\theta}.
\end{equation}
As for the incorporation of behavior information, we could still follow the similar expansion:
\begin{equation}
    f'(q,a,m) = R'_f(q,a,m)e^{i\Theta'_f(q,a,m)}.
\end{equation}

The magnitude $R_f(\cdot)$ is independent of position $m$, thar is: $R_f(q,m) = \vert\vert q \vert\vert$, which is only related to the query $q$ itself. Since behaviors are the interaction modes between users and items, from the perspective of determined sequences, behaviors only serve as attributes corresponding to tokens at each position in the sequences, thus they are also independent of position. Therefore, it can be obtained:
\begin{equation}
    R'_f(q,a,m) = \vert\vert r_f(q, a) \vert\vert,
\end{equation}
where the function of $r_f(\cdot)$ is to inject behavior information into the token, to change the magnitude of the vector to make it behaviorally relevant. We can set it through a fit for behavior representation $h(\cdot)$ as: $r_f(q, a) = h(a)q$.

The argument $\Theta_f(\cdot)$ is a function that is only dependent of position $m$ and independent of token information $q$, used to ensure the relativity of positions, that is: $\Theta_f(q,m) = \vartheta(m)$. Considering the asymmetric deterministic between items and behaviors, under determined circumstances, behaviors are heavily dependent on items, therefore ignoring item information also leads to not needing to consider behavior information. Then we can obtain:
\begin{equation}
    \Theta'_f(q,a,m) = \Theta_f(q,m) = \vartheta(m).
\end{equation}

From the above, we can conclude BaRoPE like Eq.\ref{eq:barope}:
\begin{equation}
    f'(q,a,m) = h(a)f(q,m) = h(a)qe^{im\theta},
\end{equation}
\begin{equation}
    f'(k,b,n) = h(b)f(k,n) = h(b)ke^{in\theta}.
\end{equation}
Meanwhile, relativity is also realized simultaneously:
\begin{equation}
\begin{aligned}
    g'(q,k,a,b,m-n) &= h(a)h(b)g(q,k,m-n) \\
    &= h(a)h(b)f(q,m)f(k,n) \\
    &= \langle f'(q,a,m), f'(k,b,n) \rangle,
\end{aligned}
\end{equation}
enabling the effective integration of behavior information into sequence learning.

\section{Baselines Introduction}    \label{app:baseline}

We compare our model with the state-of-the-art models from three lines of research topics, with the brief introduction as following:

\noindent \textbf{Single-Behavior Sequential Models:}
\begin{itemize}
    \item \textbf{BERT4Rec}~\cite{bert4rec}\textbf{:} A model employs bidirectional modeling of sequential information based Transformer, while being trained and optimized through Cloze task.
    \item \textbf{DiffuRec}~\cite{diffurec}\textbf{:} A framework utilizes diffusion model to probabilistically model user interests and reconstructs the target item representation through reverse process.
\end{itemize}
\noindent \textbf{Multi-Behavior Non-Sequential Models:}
\begin{itemize}
    \item \textbf{MBGCN}~\cite{mbgcn}\textbf{:} A model leverages two propagation methods to learn the embeddings of users and items to capture the different semantics of behaviors.
    \item \textbf{MB-GMN}~\cite{mb-gmn}\textbf{:} A framework explores correlations of multiple types of user behavior into meta-knowledge network.
\end{itemize}
\noindent \textbf{Multi-Behavior Sequential Models:}
\begin{itemize}
    \item \textbf{MB-STR}~\cite{mb-str}\textbf{:} A framework based transformer explicitly models the dependencies among multiple behaviors.
    \item \textbf{PBAT}~\cite{pbat}\textbf{:} A model employs personalized behavior pattern generator and behavior-aware collaboration extractor for the modeling of user preferences. 
    \item \textbf{MISSL}~\cite{missl}\textbf{:} A approach unifies multi-behavior and multi-interest modeling to obtain user profiles and utilizes self-supervised learning to refine interest representations.
    \item \textbf{M-GPT}~\cite{m-gpt}\textbf{:} A method learns dynamic behavior-aware multi-grained preferences by capturing interaction dependencies from multiple levels.
\end{itemize}

\section{Time and Space Analysis} \label{app:complexity}

\begin{figure}[H]
    \vspace{-0.3cm}
    \centering
    \subfloat[Yelp]{
        \includegraphics[width=3cm]{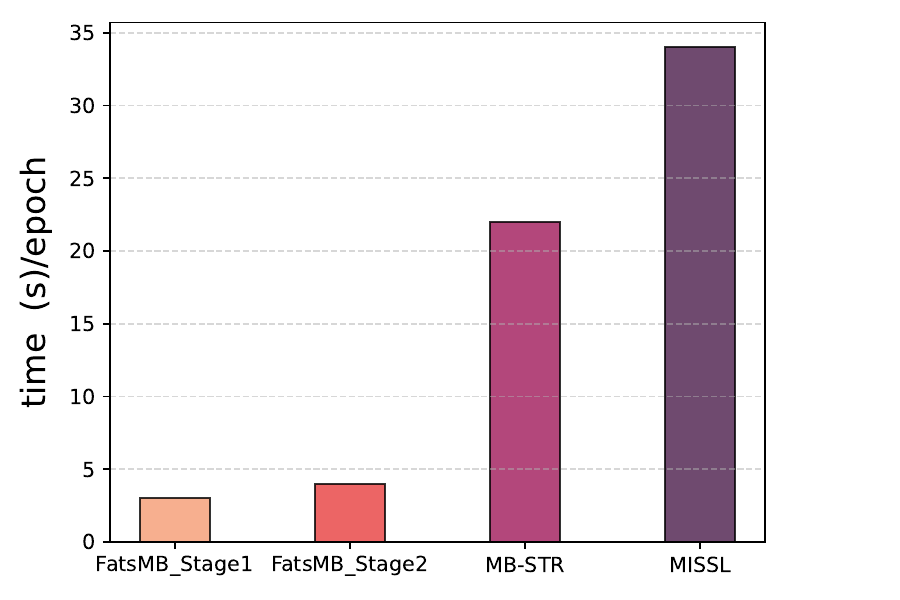}
    }
    \hspace{-0.6cm}
    \subfloat[Retail]{
        \includegraphics[width=3cm]{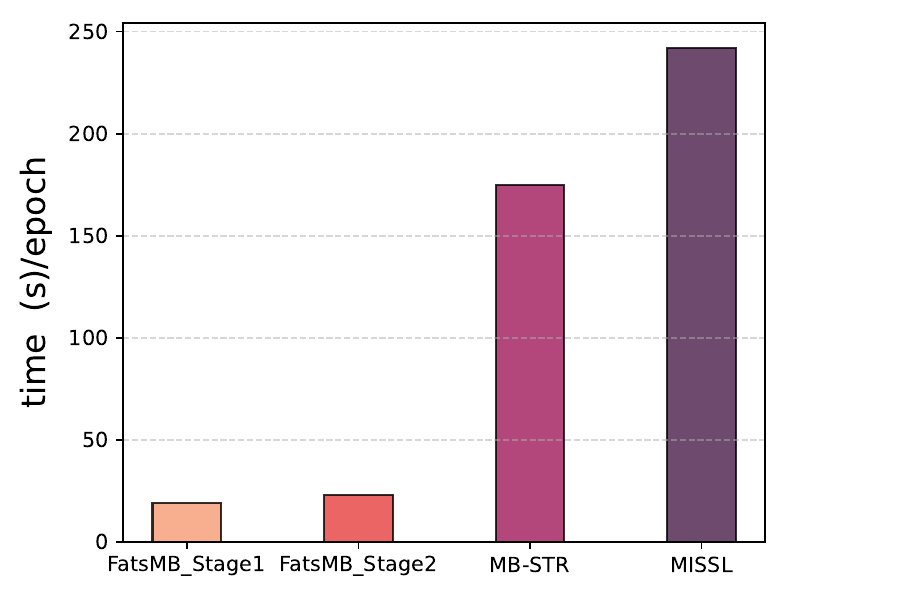}
    }
    \hspace{-0.6cm}
    \subfloat[IJCAI]{
        \includegraphics[width=3cm]{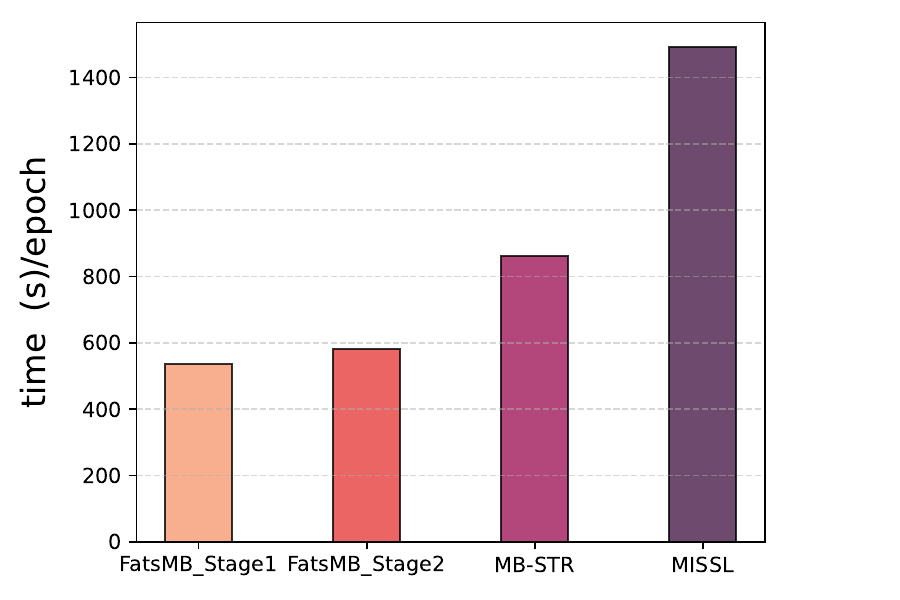}
    }
    \vspace{-0.3cm}
    \caption{Comparison of computational time.}
    \vspace{-0.3cm}
    \label{fig:time}
\end{figure}

\begin{figure}[H]
    \vspace{-0.3cm}
    \centering
    \subfloat[Yelp]{
        \includegraphics[width=3cm]{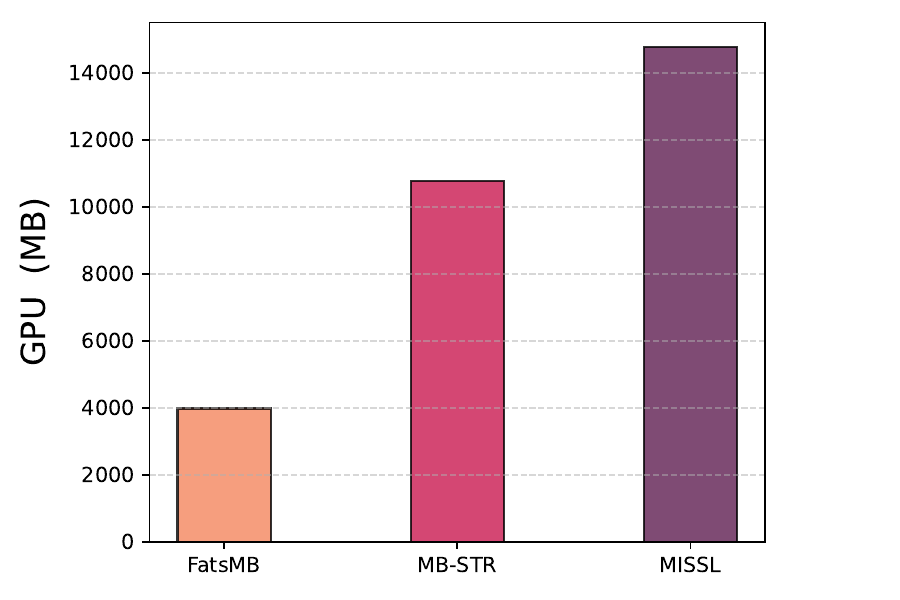}
    }
    \hspace{-0.6cm}
    \subfloat[Retail]{
        \includegraphics[width=3cm]{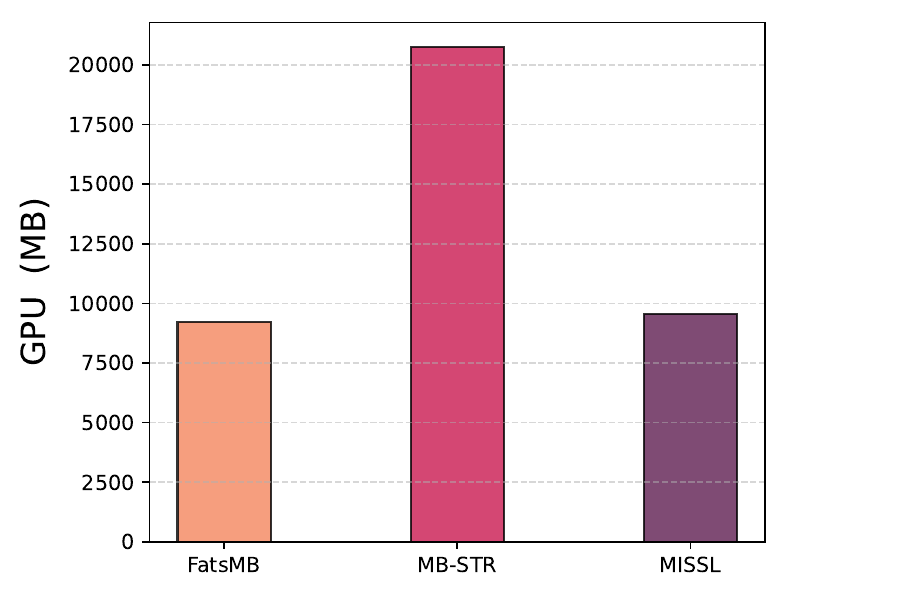}
    }
    \hspace{-0.6cm}
    \subfloat[IJCAI]{
        \includegraphics[width=3cm]{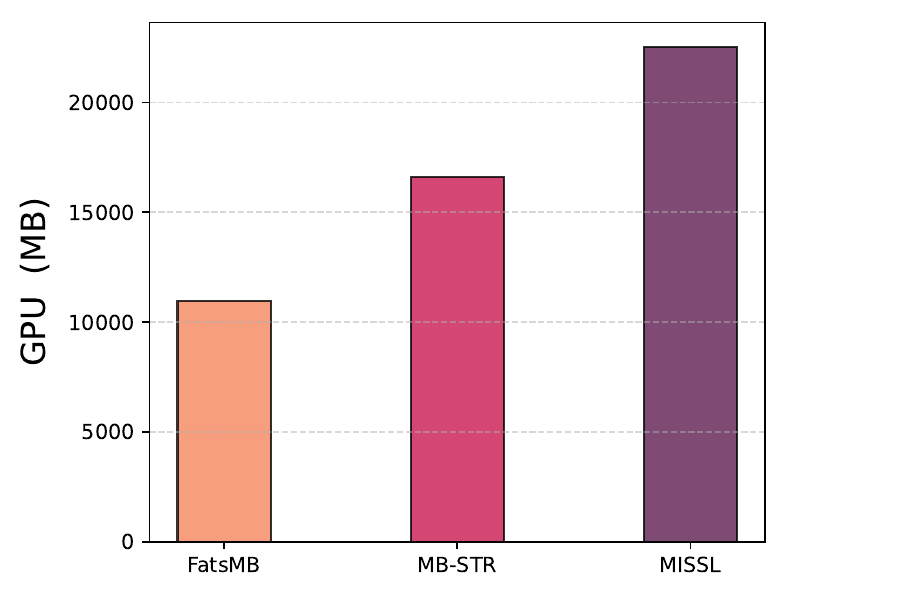}
    }
    \vspace{-0.3cm}
    \caption{Comparison of computational space.}
    \vspace{-0.1cm}
    \label{fig:gpu}
\end{figure}

We conduct comparative experiments of the time and space required for model computation. The results are shown in Figure \ref{fig:time} and Figure \ref{fig:gpu}, with all values obtained from running on a single GPU and the same hyperparameter settings. The time cost of FatsMB is reported separately by training stage, while the space cost is reported only the maximum usage due to unified model loading. We can find that both time and space cost of FatsMB are significantly reduced compared to baselines. Considering the performance advantage, we believe our model is efficient.

\end{document}